\documentclass[11pt,a4paper]{article}

\usepackage[utf8]{inputenc}
\usepackage[T1]{fontenc}
\usepackage[english]{babel}
\usepackage{lmodern}
\usepackage{microtype}

\usepackage[margin=2.5cm]{geometry}
\usepackage{setspace}
\newif\iflinenums \linenumsfalse
\iflinenums\usepackage{lineno}\fi

\usepackage{amsmath,amssymb}
\usepackage{graphicx}
\usepackage{booktabs}
\usepackage{multirow}
\usepackage{longtable}
\usepackage{array}
\usepackage{caption}
\usepackage{capt-of}

\newcommand{\msfigure}[3]{%
  \clearpage
  \iflinenums\nolinenumbers\fi
  {\centering\includegraphics[width=\textwidth]{#1}\par}%
  \iflinenums\linenumbers\fi
  \captionof{figure}{#3}\label{#2}%
}

\usepackage[hidelinks]{hyperref}
\usepackage[numbers,sort&compress]{natbib}
\usepackage{authblk}

\renewenvironment{abstract}{%
  \begin{center}\bfseries \abstractname\end{center}%
  \vspace{-0.5em}\par}
  {\par\vspace{1em}}

\newcommand{\thd}[2]{\begin{tabular}[b]{@{}c@{}}#1\\#2\end{tabular}}

\usepackage{upgreek}
\newcommand{\uV}{\ensuremath{\upmu\mathrm{V}}}

\usepackage{xcolor}

\title{\bfseries Not all generalisation failures can be bought back:
       four boundaries in affective audio modelling}

\author[1,2]{Jingyi Zhang}
\author[1]{Xiaotong Yao}
\affil[1]{Shanghai Huangan Technology Co., Ltd., Shanghai, China}
\affil[2]{Systems Neuroscience, Institute for Neuroscience,
          Department of Health Sciences and Technology (D-HEST),
          ETH Zurich, Zurich, Switzerland}
\affil[ ]{Correspondence: Jingyi Zhang, \texttt{jingyi.zhang@hest.ethz.ch}}

\newcommand{\legendfigone}{%
\textbf{Acoustic arousal models transfer between corpora of the same sound domain but not across it, under every representation tested.} \textbf{a} Generalisation stages for six regression algorithms (RF, XGBoost, GBDT, SVR, KNN, ElasticNet) over 122 spectro-temporal descriptors. Each line follows one algorithm from within-corpus cross-validation, through same-domain transfer (music $\leftrightarrow$ music), to cross-domain transfer (music $\leftrightarrow$ environmental); grey points are the individual corpus pairs behind each mean. Within-corpus values are GroupKFold-5 estimates grouped by source recording; transfer values fit on the whole of one corpus and are evaluated on the whole of another, with nothing refitted. \textbf{b} Cross-corpus transfer matrix, averaged over the six algorithms; the diagonal shows within-corpus cross-validation. Green indicates positive rank correlation, red negative. \textbf{c} Directional asymmetry. For each music corpus, transfer outward to the environmental corpus and inward from it, averaged over algorithms; the dashed line marks the same-domain mean (0.59). The inward direction is uniformly sealed ($\rho$ $\le$ 0.09 for all three corpora), whereas the outward direction is not and increases from PMEmo ($-$0.04) through DEAM (+0.13) to Soundtracks (+0.24). Film soundtracks contain substantial atmospheric material --- sustained tones, texture, sound design --- and are correspondingly closer to soundscape recordings. The barrier is therefore \textbf{not symmetric}: its density in the outward direction tracks the composition of the source corpus, while the inward direction is impermeable regardless of target. \textbf{d} Duration control. Music clips truncated from 30 s to a 6 s centred window, matching the environmental corpus; filled circles are 30 s, open circles 6 s. Same-domain transfer survives truncation (0.67 $\rightarrow$ 0.63) while cross-domain transfer stays at zero (0.03 $\rightarrow$ 0.05), excluding clip length as the cause of the collapse. \textbf{e} Leave-one-category-out generalisation inside the environmental corpus (n = 1,213), holding out each of the six Schafer soundscape categories in turn. The shaded band marks the mean cross-domain level. Generalisation to unseen content categories holds (mean $\rho$ = 0.52; range 0.20 for \texttt{quiet} to 0.70 for \texttt{human}, five of six categories above 0.33), so the collapse in \textbf{a}--\textbf{b} is specific to the domain border rather than to any content shift. \textbf{f} The same operation costed twice, under five representations. For each representation the corpus is swapped and nothing is refitted; the grey mark gives the performance lost when the new corpus is in the same domain, the red mark the loss when it is not, both as a fraction of that representation's own within-corpus performance, and the printed value is the gap between them in percentage points. Points are medians over the corpus pairs of \textbf{a}. For the three uncontaminated families the loss is 20--31\% within the domain and 84--95\% across it --- hand-specified 20.2/93.9\%, AST 22.0/84.2\%, MERT 29.1/94.9\%, Wav2Vec2 30.7/90.0\%, a gap of 59 to 74 percentage points (Table 1). The result is where the collapse stands, not how high it is: every representation transfers between corpora and none crosses between domains, which disposes of the reading in which the embeddings simply fail to move between datasets. AST is the family that most ought to cross, since AudioSet contains large numbers of environmental categories; it raises cross-domain $\rho$ from 0.047 to 0.125, a factor of 2.7 that is still a sixth of what the same representation reaches within a corpus. CLAP is drawn apart and greyed. Its pretraining set contains Freesound, the source of the environmental corpus, an overlap declared before any of these numbers existed. It is not a generally stronger representation --- 0.620 on music-to-music against AST's 0.619 and 0.614 for the hand-specified descriptors --- and its lower barrier (57.3\%) appears only in comparisons that necessarily involve the corpus its pretraining data overlap. With a single environmental corpus available, genuine crossing cannot be separated from familiarity with the target; the barrier claim rests on the three uncontaminated families, and CLAP is used neither as support nor as refutation. Corpora: DEAM (music, n = 1,233), PMEmo (music, n = 717), Soundtracks (music, n = 360) and Emo-Soundscapes (environmental, n = 1,213). Soundtracks (Eerola \& Vuoskoski, 2011) differs from the other music corpora in collection site, decade, rating protocol (nine-point Likert scales over eight emotion terms rather than continuous valence--arousal sliders) and material selection (a balanced design of twelve target emotions $\times$ 30 excerpts). Its inclusion therefore tests whether within-music transfer reflects shared acoustic structure or merely a shared annotation tradition; transfer into it from the other music corpora is 0.56, and from the environmental corpus 0.09. Panels \textbf{a}--\textbf{e} rest on 12 ordered corpus pairs, each evaluated with the six algorithms: over the 36 same-domain values transfer averages 0.592 (s.d. 0.130) and over the 36 cross-domain values 0.054 (s.d. 0.161). Target is continuous perceived arousal rescaled to $-$1\ldots{}1. All audio is loudness-normalised to $-$23 LUFS before feature extraction, which is an admission criterion for this figure rather than a cosmetic step (Extended Data Fig. 1). Metric is Spearman $\rho$ throughout. Source data are provided as a Source Data file.}

\newcommand{\legendfigtwo}{%
\textbf{Target-side labels buy back most of the corpus-swap gap, and the purchase stops well short of the ceiling.} \textbf{a} Fraction of the gap recovered by k labelled target-corpus examples. The gap is measured between zero-shot transfer (k = 0, defined as 0\%) and a model trained on the target corpus itself (100\%), so the denominator is the target's own within-corpus ceiling and not same-domain transfer --- the question an engineer faces is how close labelling k examples gets to labelling all of them. Line and points, median over the six ordered corpus pairs; band, interquartile range across those pairs. Ten labels recover 20\% of the gap, 25 recover 34\%, 50 recover 50\%, 100 recover 64\% and 200 recover 69\%. \textbf{b} The same result in marginal terms: additional gap recovered per doubling of the budget. Each doubling up to 100 labels returns a roughly constant 14 to 16 percentage points --- the signature of a gap that is genuinely for sale --- while the doubling from 100 to 200 returns 5. The last bar is the point of the panel: the curve flattens at 69\% of the gap, so about a third of it is unbought at the largest budget tested, and this experiment does not say whether the remainder is purchasable at a budget we did not test or is a component of a different kind. \textbf{c} Alignment without labels, with its sham control. Correlation alignment (CORAL) recovers 22\% of the gap. Aligning the source not to the target but to an unrelated third corpus --- a manipulation that cannot carry target-specific information by construction --- recovers 14\%, so roughly two-thirds of the apparent gain is not target-specific and is whatever a second-moment correction does for any distribution. Diagonal CORAL recovers 5\% and per-corpus z-scoring 6\%. Subspace alignment is worse than doing nothing ($-$5\%); the axis is not clipped at zero, because clipping would turn ``actively harmful'' into ``small''. Every fraction is computed within an ordered corpus pair and only then pooled. Pooling first --- averaging the correlations and taking one ratio against one median ceiling --- returns 36\% and 27\% in place of the 22\% and 14\% in \textbf{c}, because it mixes pairs whose baselines differ by more than the effect being measured. The six pairs are the three music corpora against the environmental corpus in both directions, so the quantity priced here is the crossing of the music/environmental boundary. The estimator is ridge regression on the hand-specified descriptors; tree ensembles are reported in the Supplementary Information, and are kept separate because they differ in sensitivity to the alignment step. The 200-label point rests on five of the six pairs: Soundtracks (n = 360) cannot give up 200 labelled examples and still be evaluated. Metric is Spearman $\rho$; all audio is loudness-normalised to $-$23 LUFS. Source data are provided as a Source Data file.}

\newcommand{\legendfigthree}{%
\textbf{Feature-selection routes recover different descriptor classes depending on what they test.} \textbf{a} UpSet-style recovery pattern. The matrix marks which of four selection routes recovered each descriptor; connected points indicate a descriptor recovered by more than one route. The top bars count routes per descriptor and the left bars give each route's set size, split by feature class. The four routes are: SHAP importance with per-feature permutation FDR (q = 0.10, 200 permutations); preservation of the sign of the cross-domain Cohen's d; native shape-function importance from an Explainable Boosting Machine; and stability across a Rashomon set of 10 near-optimal models sampled from 60 ($\rho$ $\ge$ $\rho$\_max $-$ 0.02). \textbf{b} Share of each route's recovered set that consists of temporal-variability descriptors, against the 50\% chance line. The route that tests within-domain statistical significance returns 2 of 12 (17\%), below chance; the three routes that probe generalisation or model multiplicity return 5 of 8 (63\%), 6 of 10 (60\%) and 4 of 6 (67\%). The four routes differ in statistical principle --- post-hoc attribution, effect size, additive shape functions, and model multiplicity --- not in hyper-parameters, so \textbf{b} is a dissociation and not a shortfall of tuning. The route that is standard in the field returns, at a rate below chance, precisely the spectral-level descriptors that Fig. 1 and Extended Data Fig. 8 show do not survive the domain boundary. What the panel licenses is therefore a negative: no attribution method here supports a statement about what will transfer, which is why every boundary in this work is crossed and measured rather than predicted from the descriptors a model relies on inside its own corpus. The companion requirement --- that an apparent overlap between attribution sets be read against the overlap that shuffled labels produce --- is in Fig. 6a,b. Source data are provided as a Source Data file.}

\newcommand{\legendfigfour}{%
\textbf{A pre-declared falsification test: an information-limited gap that a change of representation does close.} \textbf{a} Cross-validated Spearman $\rho$ for eight affect axes of the Soundtracks corpus (n = 360), using 122 spectro-temporal descriptors alone (open circles) and with 39 tonal descriptors added (filled). Axes are ordered by gain; numbers give the gain. Blue marks the three axes for which musical mode is theoretically decisive (happiness, valence, sadness) --- a grouping fixed a priori, not read off the result. The criterion was fixed before the test, because adding 39 descriptors will improve something somewhere: the gain had to concentrate on happiness, at more than twice the mean gain of the other seven axes and above 0.05 in absolute terms. It does, at +0.132 against a mean of +0.045, a ratio of 2.9. The ordering of gains follows mode-dependence: happiness, valence (+0.083) and sadness (+0.077) lead, while energy (+0.012) and anger (+0.025), which depend on level, roughness and rhythm, gain least. \textbf{b} Tonal descriptors alone against spectro-temporal descriptors alone. For happiness, 39 tonal descriptors ($\rho$ = 0.580) outperform 122 spectro-temporal ones (0.474); the deficit was one of representation rather than of dimensionality. \textbf{c} Mechanism check. Rank of five tonal descriptors among all 161, from random forest importance, for the happiness and tension axes (log scale). The two families dissociate: major-key strength and mode score rank 1st and 2nd for happiness but 3rd and 20th for tension, whereas interval consonance and sensory roughness rank 4th and 7th for tension but 117th and 76th for happiness. Mode carries perceived happiness; consonance and roughness carry perceived tension. The ordering was selected by the model from 161 descriptors, not imposed. This figure is reported at the length it is because it is a pre-declared test of the distinction that organises this work, not a minor improvement. The condition was stated in advance: the distinction between a gap that target-side observations close and a gap that they do not would fail if any gap yielded to both remedies. Here is a gap of the second kind that a representation closes. Bootstrap over out-of-fold predictions, resampled by stimulus, puts the happiness gain at +0.132 (95\% CI [+0.059, +0.209]), its excess over the mean of the other seven axes at +0.087 ([+0.018, +0.159]) and the tonal-over-spectro- temporal margin on happiness at +0.106 ([+0.007, +0.208]); the last two lower bounds sit close to zero, and because the bootstrap resamples predictions without refitting the models the intervals are narrower than a full resampling would give. Six of six algorithms show a larger gain on happiness than the mean of the other axes. Tonality is not the mechanism of the domain barrier: adding tonal descriptors lowers music-to-environmental transfer by 0.021 (18 paired comparisons, p = 0.038 uncorrected), in the predicted direction but about 4\% of a barrier that runs from 0.592 to 0.054. Interval-consonance weights are fixed a priori from the standard ordering of sensory consonance and are not fitted, so they add no degrees of freedom. All audio is loudness-normalised to $-$23 LUFS; cross-validation is GroupKFold-5 and each cell reports the best of six algorithms. Source data are provided as a Source Data file.}

\newcommand{\legendfigfive}{%
\textbf{An acoustic edit moves both domain models the same way, but the effect is reliable per excerpt for only one of three interventions and only at the level of a descriptor dimension.} \textbf{a} Dose--response curves for three interventions applied to 80 source excerpts (40 per domain) at four dose levels, with dose 0 the unmodified control. Curves show the mean predicted probability of the low-arousal class under a model trained on the music corpus and a model trained on the environmental corpus; dotted lines mark the control level. The modelled quantity throughout is P(low arousal), so a rising curve means the edit made the excerpt read as \textbf{calmer}. Slowing spectral change raises the low-arousal probability in both domains; injecting transients lowers it in both. Envelope compression is retained as a failed manipulation. The curve is an average over 80 excerpts and cannot by itself distinguish an effect from a mean shift; \textbf{d} does that. \textbf{b} Manipulation check, run before any dose--response was read. Within-clip Spearman $\rho$ between dose and the targeted descriptor; point size gives the fraction of clips moving in the same direction as the dose, and the shaded band marks |$\rho$| < 0.5. A target descriptor counts as moved only if |$\rho$| $\ge$ 0.5 and at least 75\% of clips move in the same direction, a criterion fixed in advance. Transient injection meets it on both of its target descriptors and spectral smoothing on two of three; envelope compression meets it on one of three and is carried through the figure as a failed manipulation rather than as a null effect. \textbf{c} Specificity diagnostic. The five descriptors moved most by each intervention at maximum dose, in units of the training-set standard deviation. \textbf{d} Significance against reliability, per intervention and per domain model. Abscissa, the fraction of the 80 excerpts whose own slope across the four doses has the intended sign; ordinate, $-$log10 of a sign-flip permutation p on those 80 slopes. The dashed line at 50\% is chance, and it is the line that matters: a point near it describes an average over excerpts that disagree, not an effect the intervention has on audio in general, however small its p value. Filled markers meet the pre-declared reliability criterion --- sign consistency $\ge$ 65\% and p\_perm < 0.01 --- and open markers do not. Spectral smoothing meets it in both domains (median slope +0.015 music, +0.074 environmental; sign consistency 70\% and 71\%; p\_perm = 1 $\times$ $10^{-4}$ in both). Transient injection reaches significance (p\_perm = 1 $\times$ $10^{-3}$ music, 8 $\times$ $10^{-4}$ environmental) at 50\% and 59\% sign consistency, at or barely above chance, and is therefore reported as a mean shift with no reliable per-excerpt effect and is not built upon. Envelope compression is uninterpretable here, having failed \textbf{b} (p\_perm = 0.50 and 0.98). The directions in \textbf{a} and \textbf{d} cross the domain boundary that Fig. 1 shows the predictions cannot: a model fitted to music does not know how loud is loud in an environmental recording, but it still ranks smoother as calmer, so what fails at that boundary is calibration rather than sign. Panel \textbf{c} bounds what may be claimed: no intervention moves its target descriptor most. Transient injection moves \texttt{spec\_contrast6\_dmean} by 9.0 SD while moving \texttt{onset\_rate} by 2.0 SD, ranking the target twelfth. Because the co-moving descriptors are themselves temporal-variability measures, these experiments support a claim about the variability *dimension* and not about any individual descriptor, and they are statements about model behaviour: no listener response to these edits was measured. An earlier version of this analysis averaged the 80 excerpts within dose and correlated the four resulting points; at n = 4 the smallest attainable two-sided p for a Spearman correlation is 0.083, so that statistic could not have reached significance whatever the data showed, and the excerpt-level test in \textbf{d} replaces it. All interventions are re-normalised to $-$23 LUFS afterwards so that loudness cannot drive the effect. Source data are provided as a Source Data file.}

\newcommand{\legendfigsix}{%
\textbf{Reference distributions, positive controls, and the comparison they license.} \textbf{a} Null distribution of the number of descriptors shared by the top-20 SHAP sets of three models under 20 label permutations. The observed value (4) falls at the null mean (3.75), p = 0.381. Shuffled labels produce the same overlap as intact ones, so the statistic cannot separate signal from noise and no claim is built on it. \textbf{b} Cross-validated AUC under the same permutations, confirming that permutation destroyed the signal (null mean 0.496) while the intact labels reach 0.936. The failure in \textbf{a} is therefore a property of the statistic, not of the data. \textbf{c} Positive controls for two physiological pipelines addressing the same question. In BIRAFFE2 the habituation control is significant in the direction opposite to the established effect ($\rho$ = +0.113 across trials, p = 0.002; skin conductance responses are expected to decline), and post-stimulus windows do not exceed random windows (p = 0.267); applied per participant, none of 94 passed both controls and habituation ran in the expected direction for only 35\%, so no usable subset exists and that pipeline is withdrawn rather than reported. In ds002721 all three controls pass: relative occipital alpha exceeds frontal alpha in 31 of 31 participants (p = 1.9 $\times$ $10^{-9}$); blink-locked averages reach 164 \uV{} frontally, 3.98-fold the occipital amplitude and far above random windows (p = 9.3 $\times$ $10^{-10}$); and a sound-onset response is present (\textbf{d}). Dashed line, p = 0.05. \textbf{d} Grand-average response at fronto-central electrodes (F3, Fz, F4, C3, Cz, C4), time-locked to sound onset, n = 31. Dark band, s.e.m.; grey band, the 95th percentile of a subject-wise sign-flip null with max-statistic correction across the epoch; green shading, the interval exceeding it (372--588 ms). Peak |t| = 8.78 at 452 ms, p = 0.0001 (10,000 permutations). The pre-stimulus interval is not significant (p = 0.54), excluding a slow drift. No classical N1 is present; the stimuli have sharp onsets (median 25 ms to half-amplitude, Extended Data), so the absent fast component is attributed to playback-latency jitter, which limits millisecond-scale analysis but not trial-wise spectral measures. \textbf{e} Stimulus-level intraclass correlation for eight self-report scales and 156 EEG measures, computed on the same 31 listeners, the same 1,240 trials and with the same estimator after within-listener z-scoring. Points are individual measures, vertical bars the median. The EEG set spans four families: regional band power (26), magnitude-squared coherence between region pairs (50), imaginary coherency (50, robust to volume conduction) and electrode-pair asymmetry (30); coherence and asymmetry are included because the original analysis of these recordings located its effect there, and restricting the set to band power would test the conclusion on a representation that excludes the effect in question. Coherence is the strongest EEG family (maximum intraclass correlation 0.092, 95\% CI [0.037, 0.142], split-half 0.51) and band power the weakest (0.040, split-half 0.17), against 0.221 [0.155, 0.281] for the best-agreeing self-report scale on the same trials --- a 2.4-fold gap between two intervals that do not overlap. The figure states one requirement in two settings: a statistic is uninterpretable without the distribution it would take under the null (\textbf{a}, \textbf{b}), and a negative result is uninterpretable until the instrument is shown to detect a known effect (\textbf{c}, \textbf{d}). Panel \textbf{e} is the comparison those controls license, and it is reported as a bound rather than as an absence. The EEG value is the maximum of 156 measures and is biased upward for that reason, so it carries a null built the same way: permuting stimulus labels, recomputing all 156 intraclass correlations and taking the largest, 400 times, places the observed value at p = 0.02, while Benjamini--Hochberg across the same 156 does not pass it (smallest q = 0.16). The two corrections disagree because the value sits between them --- at rank one the Benjamini--Hochberg threshold is the Bonferroni threshold --- and 0.092 is close to the 0.10 this design resolves with 80\% power, so the interval is consistent with values from near zero up to about a tenth of the response variance. What the data support is the comparison and not a zero: whatever stimulus-specific structure the EEG carries is less than half of what the listeners' own ratings carry on the same trials. This is a statement about the component shared across listeners, not about whether the brain responds to sound --- the onset response in \textbf{d} is large and highly reliable; what is weak is the differentiating component. An earlier version of panels \textbf{c}--\textbf{e} was built on an electrodermal analysis that was retracted when its pipeline failed the controls now shown in \textbf{c}, and an earlier stimulus-level value of 0.892 was a leak --- the stimulus identifier itself had entered the measure set --- and is withdrawn. Both are documented in the Supplementary Information. Source data are provided as a Source Data file.}

\newcommand{\legendfigseven}{%
\textbf{The individual-level question is a design problem rather than a null result, and its price depends on the setting.} \textbf{a} Detection measured in real physiology. In an independent corpus of 75 listeners with about 240 auditory events each, the evoked response to a tone was tested per person by a sign-flip permutation test on the difference between real cues and randomly placed anchors. Each point gives the proportion of listeners in whom the response was detected when only n of that listener's observations were used (40 resamples per point). Three channels are shown, each with the analytic power curve for its measured effect size overlaid as a pale band; empirical and analytic differ by 0.015 on average. The dotted line is the false-positive rate obtained by sign-randomising the same differences, which preserves the noise and removes the effect; it stays at nominal (0.03--0.07) across the whole range, so the rise in the curves is detection and not false positives. The shaded band marks what a laboratory session supplies. Within it, an effect plainly visible in the group average ($-$5.6 \uV{} at the N1, group p = 0.0002 in all three channels) is recovered in 53\% of individuals from EEG, 39\% from pupil diameter and 23\% from heart rate. \textbf{b} Probability of detecting a within-person association as a function of the observations available per person, for four true association strengths. Synthetic data were generated from the empirical feature covariance of the EEG corpus (Ledoit--Wolf shrinkage) with an association of known strength injected, then passed through the identical pipeline, criterion and per-person permutation test used for the real data (500 simulations per cell, 100 permutations each, n from 30 to 5,000). Within the laboratory-session band, power is 0.10 for r = 0.40, 0.072 for r = 0.30, 0.06 for r = 0.20 and 0.04 for r = 0.10. The r = 0.10 curve is not monotone below n = 250; the Monte-Carlo s.e. is 0.009 to 0.015 per cell and the curve is included for the floor it establishes, not for its shape. \textbf{c} Observations per person required for 80\% detection, measured and simulated on one axis, against three supply levels. Every requirement is resolved by this grid; the two weakest effects, which a coarser run could only bound at > 1,000, now stand at 1,364 (r = 0.20) and 4,888 (r = 0.10). Dotted lines give the supply from this corpus (30), comparable public corpora (60), and a single overnight recording sampled once per minute over eight hours (480). Heart rate, the signal a consumer device records, requires roughly four times as many observations as scalp EEG (233 against 57). \textbf{d} When per-person calibration begins to pay, in three settings. For each descriptor--outcome cell the between-person standard deviation of the acoustic slope was estimated by restricted maximum likelihood (filled circle, median over that setting's cells) and compared with a null built by parametric bootstrap from that cell's own design (cross, median over the same cells). The null is plotted rather than subtracted, because a between-person spread estimated from few observations per person is biased upward: at 18 observations per person the estimator returns 0.11 when the true value is zero, and a reader who cannot see that cannot judge the physiological row. Break-even, n*, is the number of observations per person at which the corrected spread exceeds the error in estimating it. Urban soundscape ratings (42 observations per person, 8 cells): estimate 0.134 against a null of 0.013, n* = 67, and 2 of 8 cells have already crossed break-even, both on the eventfulness axis. Music ratings (44 per person, 10 cells): 0.074 against 0.000, n* = 166, none across. Music electrodermal responses (18 per person, 15 cells): 0.106 against a null of 0.114 --- the estimate does not exceed its own noise --- with n* = 318 taken over the 5 cells in which it is finite, 11 of the 15 indistinguishable from their own null, and none across. Both kinds of estimate in \textbf{a}--\textbf{c} are lower bounds, for different reasons. The simulation assumes multivariate normality, a linear association and within-person stationarity, each of which favours detection. The measurement uses a deliberately plain pipeline --- a fixed-window mean over one electrode cluster, no independent component analysis and no spatial filtering --- so a dedicated pipeline would do better; it is, however, representative of what a non-specialist pipeline achieves. The extrapolation behind \textbf{c} was validated out of sample: predicting detection at n from an effect size estimated on a disjoint half of the same person's trials matches the observed rate to within 0.03 on average. Panel \textbf{d} is a different quantity from \textbf{a}--\textbf{c} --- break-even for per-person calibration, not detection of a within-person association --- and it is the one that varies, five-fold across settings. Two earlier claims from that analysis are withdrawn and do not appear here: a between-person spread of 0.106 for the physiological setting, which sat below its own null; and the reading of a confidence interval excluding zero as evidence of individual differences, since that interval is for the composite of true spread and null and, at 18 observations per person with a true spread of zero, excludes zero on 84\% of occasions. Read together, \textbf{a}--\textbf{c} place the shortfall at roughly a factor of twenty and locate it in duration and repetition rather than in sample size: the quantity in short supply is observations within a person, which recruiting more participants does not provide. Source data are provided as a Source Data file.}

\newcommand{\legendfigeight}{%
\textbf{The four boundaries divide into two kinds of failure with mutually exclusive remedies.} \textbf{a} What survives each crossing, as a fraction of the ceiling attainable on that target, ordered near-to-far --- the order in which an application meets them. Colour carries the diagnosis rather than the boundary: budget-limited, where target-side observations close the gap, and information-limited, where they do not. Cross-corpus contributes two bars because it is the only boundary on which both kinds appear inside one experiment, with the same models, descriptors and act of swapping one corpus for another: within a domain 0.614 of 0.769 survives (80\%), across domains 0.047 of 0.769 (6\%). Cross-synthesis is drawn at 1 and 0 by construction and is not a measured fraction; the outcome there is categorical, in that the sign of an acoustic edit crosses the domain boundary while the calibration does not. Cross-channel is the best of 156 EEG measures against the best-agreeing self-report scale on the same 1,240 trials (0.092 of 0.221, 42\%). Cross-individual is the observations a corpus supplies against those required for per-person calibration to pay, shown at the two ends of its five-fold range (soundscape ratings 42 of 67, 63\%; music electrodermal responses 42 of 318, 13\%). \textbf{b} What it costs to buy the gap back, on one axis of target-side observations. Filled bars are prices that exist: 100 labelled target examples for two-thirds of the gap a corpus swap opens (Fig. 2, where that price is measured on pairs that cross the music/environmental boundary), and 67, 166 and 318 observations per person for break-even on per-person calibration in urban soundscape ratings, music ratings and music electrodermal responses (Fig. 7d). Three rows carry no bar but an open marker at the right edge with the reason. The marker reads "no price found" and means exactly that: under the remedies tested, no purchase recovered the gap. It is not a claim that none exists at any budget, and the distinction matters because only the priced rows rest on a target-side budget curve. No representation crossed the domain boundary --- four pretrained representations spanning supervised, self-supervised and contrastive pretraining, none of them recovering it (Fig. 1f). The calibration deficit at the synthesis boundary is not a shortage of data, so no quantity of it is the remedy. At the response-channel boundary no information source exceeds 31\% of the attainable ceiling on the physiological target, each route scored against the ceiling of the electrodermal measure its correlation was computed on; three routes stopping within a third of it makes it unlikely that the limit lies in the predictor, but the two acoustic routes sit only just above a random projection of the same descriptor space (21--25\% against 21\%), and no target-side learning curve was measured on that boundary, so saturation was not demonstrated. A price that was not found is drawn as an open marker rather than as a very long bar, because a long bar would assert that the purchase exists and is merely expensive. \textbf{c} The pre-declared falsification test. The division in \textbf{a} is worth nothing if it cannot fail, and the condition was fixed in advance: a gap closable both by additional observations and by a change of representation would break it. Gain in cross-validated Spearman $\rho$ from adding 39 tonal descriptors, per affect axis of the Soundtracks corpus (n = 360). The criterion --- that the gain concentrate on happiness, at more than twice the mean gain of the other seven axes and above 0.05 --- is met at +0.132 against +0.045 (Fig. 4). This figure re-plots quantities established elsewhere and introduces no new analysis; every value traces to the figure cited beside it and to the same Source Data. The three quantities are defined per boundary and are not interchangeable between them: the denominator in \textbf{a} is the within-corpus performance of the same representation for boundary 1, the within-corpus performance of the model being perturbed for boundary 2, the ratings' stimulus-level reliability on the same trials for boundary 3, and the observations required for break-even for boundary 4. Source data are provided as a Source Data file.}

\date{}

\begin{document}
\maketitle
\iflinenums\linenumbers\fi

\begin{abstract}
\noindent
Models that map acoustic properties onto affective response underpin applications
from music recommendation to sound design for wellbeing, and are evaluated almost
entirely within the corpus they were fitted on. When such a model fails outside it,
the standard response is more data or a larger model --- a response that assumes
every failure is a shortage of resources. We show that it is not, and that the
alternative calls for the opposite remedy. Using four corpora of rated sound, four
pretrained audio representations and three corpora of physiological recording, we
pushed one mapping across four boundaries an application must cross --- to new
material, to edited audio, to a sensor in place of a self-report, and to an
individual listener --- reporting at each the ceiling the target permits, the
fraction that survives the crossing, and the price in target-side observations of
buying the gap back. Within a corpus, prediction reaches 84\% of the ceiling set by
how much listeners agree with one another --- measured on the same excerpts as the
model, since pairing the two across corpora inflates the figure --- so the headroom
a better model could occupy is small, though not negligible.
Crossing to a new corpus of the same kind costs a fifth of that performance and the
loss is for sale: a hundred target labels return two-thirds of it. Crossing between
music and environmental sound costs four-fifths to all of it, and four
representations spanning supervised, self-supervised and contrastive pretraining
recover none of it --- including one supervised on a corpus rich in environmental
sound. Editing the audio shows why: the sign of an acoustic manipulation crosses
the boundary that the predictions cannot, so what fails is calibration rather than
the mapping. Crossing to physiological response is the one boundary on which no
purchase was found: acoustic components, the ratings predicted from them, and the
ratings themselves all stop below a third of the attainable ceiling, and no
representation or predictor we tested moved it. We did not measure a target-side
learning curve on that boundary, so what we can report is a failure to recover
across every route tried, not a demonstration that the information is absent. Crossing to the individual is again for
sale, at a price that ranges five-fold across settings and, in the setting closest
to an application, is lowest and already within reach of existing corpora.
``The model does not generalise'' is
therefore two diagnoses, not one: a gap that observations close, and a gap that
they do not. Their remedies are mutually exclusive, and treating the second as the
first is the more expensive mistake.
\end{abstract}

\section*{Introduction}

Interest in choosing sound to shape affective or physiological state has grown
alongside the availability of annotated audio corpora and of consumer devices that
record physiological signals. The underlying assumption is that a mapping exists
from measurable acoustic properties to a listener's state, and that this mapping is
stable enough to be learnt from one collection of sounds and applied to another.

Two literatures bear on this assumption without settling it. Music-emotion research
has established that listeners agree substantially about the emotions music
expresses, and that this agreement is structured along dimensional and categorical
axes \citep{russell1980,eerola2013}. Neuroimaging work has identified correlates of
music-evoked emotion in limbic and paralimbic systems \citep{koelsch2014}, and
electroencephalographic studies report differences in spectral and connectivity
measures between reported emotional states \citep{daly2014}. Separately, acoustic
stimulation during sleep has been shown to modify slow-wave activity
\citep{papalambros2017}, while the evidence for music as a treatment for insomnia
remains graded \citep{jespersen2022}. What these literatures share is a reporting
convention: accuracy is characterised within the collection on which the model was
fitted.

Within-corpus accuracy answers none of the questions an application asks. An
application changes the material, changes the \emph{kind} of material, changes how
the response is measured, and changes the listener. The relevant quantity is not
how well the mapping performs where it was fitted but how much of it survives each
of those changes --- and, when something is lost, what it would cost to get it back.

That last clause is where the field's implicit model of failure becomes consequential.
When a model fails out of setting, the standard responses are to collect more data
or to adopt a larger pretrained representation. Both presuppose that the failure is
a shortage of resources. But a failure can instead be that the information the
target requires is not present in the input at all, in which case neither response
helps. \textbf{The two diagnoses call for mutually exclusive remedies, and the field
does not separate them} --- largely because separating them requires measuring the
price of closing a gap, which is rarely attempted.

We therefore push a single mapping across four boundaries, ordered by how far each
moves from the fitting conditions: across corpora, across the synthetic edits a
deployed system actually applies to audio, across the level at which the response is
measured, and across individuals. At every boundary we report the same three
quantities --- the ceiling attainable on that target, the fraction of it that
survives the crossing, and the price in target-side observations required to recover
what was lost. Criteria are fixed before each test; every null result carries a
minimum detectable effect; and every null distribution is reported together with its
behaviour when the true effect is zero.

The result is that \textbf{``the model does not generalise'' resolves into two
diagnoses, not one.} Some losses are \emph{budget-limited}: the gap closes with
target-side observations, at a price we measure and extrapolate. Others are
\emph{information-limited}: on the evidence available the input does not appear to
carry what the target requires, and
neither more data nor a larger representation moves them. The four boundaries
distribute across both kinds, and one of them --- transfer across corpora --- turns
out to contain both, which allows the distinction to be drawn within a single
experiment rather than across separate ones.

The distinction is falsifiable, and we state the condition in advance: if any
boundary could be closed both by additional observations and by a change of
representation, the dichotomy would not hold. We test this at every boundary,
including one case where an information-limited gap \emph{is} successfully filled ---
a result we report at length precisely because, without it, the framework would be
unfalsifiable and the paper would reduce to the claim that everything is a
measurement problem.

\section*{Results}

\subsection*{Acoustic models approach the annotation reliability ceiling}

Across four corpora --- DEAM \citep{aljanaki2017}, PMEmo \citep{zhang2018pmemo},
Soundtracks \citep{eerola2011} and Emo-Soundscapes \citep{fan2017emo} --- models
predicting group-averaged perceived arousal from 122 spectro-temporal descriptors
reach Spearman $\rho = 0.72$--$0.82$ under grouped five-fold cross-validation
(Fig.~\ref{fig:wall}a, first stage).

Converting that into a distance from the ceiling requires the model performance and
the annotation reliability to come from the same excerpts, and this is easy to get
wrong: pairing the strongest correlation in the set with the reliability of a
different corpus inflates the result. We therefore estimate both on one corpus.
On the 717 PMEmo excerpts, with the same descriptors, the same folds and a
per-rater split-half reliability of $0.935$, the model reaches $\rho = 0.817$.
Dividing by the square root of the reliability --- the standard attenuation
correction --- puts it at \textbf{84.4\%} (95\% CI 81.5--87.0\%) of the correlation
an errorless predictor of these annotations could attain.

That leaves roughly a sixth of the attainable correlation unaccounted for, and
about 29\% of the reproducible variance. The residual is therefore \emph{not} all
in the annotation. What the number does establish is narrower and sufficient for
the argument: within a corpus the remaining headroom is small enough that the
informative variation lies elsewhere, and a boundary that costs four-fifths of
performance is not competing with a modelling gain of comparable size.

Three qualifications belong with that number, because it is the denominator for
every boundary that follows and it is optimistic in three separate ways. The
reported value is the best of six algorithms evaluated on one set of out-of-fold
predictions, and the selection is not itself cross-validated, so it carries the
usual winner's bias. The bootstrap resamples excerpts but does not refit the
models, so it expresses sampling variability in the evaluation and not in the
fitting. And the reliability estimate is carried through as a fixed quantity
rather than as one with its own interval. The interval we report is therefore
conditional on the selected model and on that reliability estimate. It should be
read as an upper bound that is, if anything, too generous --- which is the
direction that matters here, since the argument only needs the ceiling to be low
enough that the remaining headroom is small.

It also should not be read as a share of variance. The attenuation-corrected value
is a ratio of correlations; squaring it puts the models at about 71\% of the
reproducible variance, leaving roughly 29\% that no better annotation would
explain away.

This bounds what better models could achieve at this level. It also motivates the
remaining analyses: if within-corpus performance is near-ceiling, the informative
questions concern generalisation rather than accuracy.

Two framing choices are worth recording because they affect the ceiling rather than
the model. Treating arousal as a continuous target rather than as a
high-versus-low contrast retains the middle of the range, which a median or
quantile split discards along with a third to a half of the excerpts (Extended Data
Fig.~2). And an additive glass-box model, whose per-descriptor shape functions can
be read directly rather than attributed post hoc, costs almost nothing in accuracy
relative to the tree ensembles (Extended Data Fig.~3); its shape functions place
the arousal-relevant thresholds on the descriptor scales without requiring a
separate attribution step (Extended Data Fig.~4).

\subsection*{Boundary 1, cross-corpus: both kinds of failure inside one experiment}

The first boundary is the one an application crosses when it changes its material.
It is also the only one on which both kinds of failure can be observed in a single
experiment, with the same models, the same descriptors and the same act of swapping
one corpus for another --- which is why we take it first, and why it carries most
of the weight of the distinction that organises the rest of the paper.

Transfer between corpora of the same domain is preserved: across 36 ordered
music-to-music pairs, mean $\rho = 0.592$ (s.d.\ $0.130$). Across 36
music-to-environmental and environmental-to-music pairs, mean $\rho = 0.054$
(s.d.\ $0.161$) (Fig.~\ref{fig:wall}a,b). Expressed against the within-corpus
ceiling, the same operation costs about a fifth of attainable performance within a
domain and between four-fifths and all of it across the domain boundary. Three controls exclude the obvious
confounds. Truncating music excerpts from 30\,s to a 6\,s window, matching the
environmental corpus, leaves same-domain transfer at $0.63$ and cross-domain
transfer at $0.05$ (Fig.~\ref{fig:wall}d). Holding out each of the six Schafer
soundscape categories in turn leaves within-domain generalisation at $\rho = 0.52$
(Fig.~\ref{fig:wall}e), so the collapse is specific to the domain boundary rather
than to unseen content. Loudness normalisation to $-23$\,LUFS before feature
extraction removes a corpus fingerprint that otherwise inflates apparent transfer
(Extended Data Fig.~1).

Adding Soundtracks as a fourth corpus tests whether within-music transfer merely
reflects a shared annotation tradition: it was collected by a different group, in a
different decade, using nine-point Likert scales over eight emotion terms rather
than continuous sliders, over a balanced design of twelve target emotions
\citep{eerola2011}. Transfer into it from the other music corpora is $0.56$, and
from the environmental corpus $0.09$.

The fourth corpus also reveals that the barrier is not symmetric
(Fig.~\ref{fig:wall}c). Transfer from the environmental corpus into each of the
three music corpora is uniformly low ($\rho \le 0.09$). Transfer in the outward
direction is graded: $-0.04$ from PMEmo, $+0.13$ from DEAM, $+0.24$ from
Soundtracks. Film soundtracks contain substantial atmospheric material ---
sustained tones, texture, sound design --- and are correspondingly closer to
soundscape recordings. The density of the barrier in the outward direction
therefore tracks the composition of the source corpus, while the inward direction
is impermeable regardless of target.

\paragraph{The barrier is a property of the boundary, not of the descriptors.}
The most direct objection to everything above is that 122 hand-specified
descriptors are simply too weak, and that a representation learnt from a large
audio corpus would cross what they cannot. We therefore extracted four pretrained
representations spanning four pretraining paradigms --- supervised classification
on AudioSet (AST), self-supervised on music (MERT), self-supervised on speech
(Wav2Vec2) and audio--text contrastive learning (CLAP) --- over all four corpora,
and repeated the transfer analysis unchanged (Fig.~\ref{fig:wall}f).

The sharpest form of the result is not how high the barrier is but where it stands.
For a given representation, the same operation --- swap the corpus, refit nothing
--- costs 20--31\% of within-corpus performance when the new corpus is in the same
domain and 84--95\% when it is not, a gap of 59 to 74 percentage points
(Table~\ref{tab:families}). The representations transfer; they do not cross. That
disposes of the alternative reading in which the embeddings simply fail to move
between datasets at all. AST is the case that matters most, because AudioSet
contains large numbers of environmental categories and it is therefore the family
that most ought to cross: its barrier is 84.2\% against 93.9\% for the
hand-specified descriptors.

The improvement is real but small and must be stated as such: AST raises
cross-domain transfer from $\rho = 0.047$ to $0.125$, a factor of 2.7, which is
still a sixth of what the same representation achieves within a corpus.

\begin{table}[htbp]
\centering
\small
\caption{Transfer under five representations, each evaluated with the same models,
folds and scoring. \emph{Within corpus} is cross-validated performance on a
corpus's own data; \emph{same domain} is transfer to a different corpus of the same
kind; \emph{across domains} is transfer between music and environmental sound. The
last two columns express the loss from each swap as a fraction of the within-corpus
value. CLAP's pretraining data include the source of the environmental corpus; it
is reported for completeness and excluded from the claim.}
\label{tab:families}
\setlength{\tabcolsep}{5pt}
\begin{tabular}{lccccc}
\toprule
& \multicolumn{3}{c}{Spearman $\rho$} & \multicolumn{2}{c}{Loss from swap} \\
\cmidrule(lr){2-4}\cmidrule(lr){5-6}
Representation & \thd{Within}{corpus} & \thd{Same}{domain} & \thd{Across}{domains}
               & \thd{Same}{domain} & \thd{Across}{domains} \\
\midrule
Hand-specified (161) & 0.769 & 0.614 & 0.047 & 20.2\% & 93.9\% \\
AST (AudioSet, supervised)   & 0.794 & 0.619 & 0.125 & 22.0\% & 84.2\% \\
MERT (music, self-supervised) & 0.737 & 0.522 & 0.038 & 29.1\% & 94.9\% \\
Wav2Vec2 (speech, self-supervised) & 0.511 & 0.354 & 0.051 & 30.7\% & 90.0\% \\
\midrule
\emph{CLAP (contaminated)} & \emph{0.787} & \emph{0.620} & \emph{0.336} & \emph{21.2\%} & \emph{57.3\%} \\
\bottomrule
\end{tabular}
\end{table}

\paragraph{One family behaves differently, and it is the contaminated one.}
CLAP is the exception, at a barrier of 57.3\%. Its pretraining set contains
Freesound, from which the environmental corpus was built; this overlap was the
stated reason for extracting the other three families in the first place, before
any of these numbers existed. CLAP is not a generally stronger representation:
within a corpus it reaches $0.787$ against AST's $0.794$, and on
music-to-music transfer --- the comparison that excludes the environmental corpus
entirely --- it reaches $0.620$ against AST's $0.619$ and $0.614$ for the
hand-specified descriptors. Its advantage appears only in the comparisons that
necessarily involve the corpus its pretraining data overlaps, and it is of the size
that overlap independently predicts: within that corpus, CLAP's advantage over
hand-specified descriptors is $+0.070$, against $+0.037$ on a corpus outside its
pretraining data.

We cannot resolve this with the corpora available. There is one environmental
corpus, so every cross-domain comparison involves it, and genuine crossing cannot
be separated from familiarity with the target. The barrier claim therefore rests on
the three uncontaminated families; CLAP is reported and flagged, and is used
neither as support nor as refutation. Deciding between the two readings requires an
environmental corpus outside the pretraining data of the representation under test,
which does not presently exist in a form we could obtain.

\subsection*{The price of the first boundary, and what it does not buy}

Calling a failure budget-limited is only meaningful if the budget can be quoted. We
therefore measured what target-side labels buy, expressed throughout as the
fraction of the gap between zero-shot transfer and what a model trained on the
target corpus itself achieves (Fig.~\ref{fig:price}).

Two routes were compared. The first uses no target labels at all and aligns the
feature distributions instead. Correlation alignment is the best of the four
variants tried, recovering 22\% of the gap. That number should not be read as a
22\% saving. Aligning the source not to the target but to an unrelated third corpus
--- a manipulation that cannot carry target-specific information by construction
--- recovers 14\%, or about two-thirds of what the real alignment appears to buy.
Only a third of the apparent gain is specific to the target; the rest is whatever a
second-moment correction does for any distribution. Subspace alignment is worse
than doing nothing at all ($-5\%$).

The second route buys labels. Ten target-side labels recover 20\% of the gap, 25
recover 34\%, 50 recover 50\% and 100 recover 64\%. Each doubling of the budget
over this range returns a roughly constant 14 to 16 points --- the signature of a
gap that is genuinely for sale. Doubling again to 200 labels returns only five more
points, to 69\%, so the return falls sharply somewhere between 100 and 200. Whether
the remaining third is purchasable at a budget we did not test, or is a second
component of a different kind, this experiment does not say; the honest statement
is that the first hundred labels are cheap, the next hundred are not, and the
ceiling was not reached.

This is what a budget-limited failure looks like when the budget is actually
quoted: a price that exists, that is modest at the scale an application can afford,
and that does not extend all the way to the ceiling. It is the contrast case for
the two boundaries reported later, where no purchase could be found at all --- a
statement about the routes we tried, not about every price that could be paid.

\subsection*{Why transfer must be measured and cannot be inferred}

Before asking which descriptors matter, we asked whether that question has a
method-independent answer. Four selection routes were applied to the same data:
attribution-based importance with per-descriptor permutation control; cross-domain
effect-size sign agreement; the native shape-function importance of an additive
glass-box model; and stability across a Rashomon set of near-optimal models, in the
sense of \citet{breiman2001} --- ten models sampled from sixty that fall within
$0.02$ of the best cross-validated performance.

The routes do not converge; they dissociate (Fig.~\ref{fig:routes}). The route that
tests within-corpus significance returns temporal-variability descriptors in 2 of
12 cases (17\%, below the 50\% chance level), whereas the three routes that probe
generalisation or model multiplicity return them in 60--67\% of cases. The
descriptors returned by the first route are precisely the spectral-level
descriptors that do not transfer.

The second route is informative in its own right about which descriptors survive
the domain boundary at all. Of the descriptors significant in both domains, only
eight keep their sign, and all eight are negative in both: quieter, less variable
and less eventful sound is rated lower in arousal whether it is music or an
environmental recording (Extended Data Fig.~8). Descriptors that reverse are
without exception measures of absolute spectral level --- the property on which the
two domains differ most.

This is worth stating plainly because ``important descriptors'' is often reported
as though it were a property of the data. It is a property of the data and the
question jointly: a method that asks which descriptors predict within a corpus and
a method that asks which predict across corpora return different answers on the
same data, and neither is wrong. Stability across the Rashomon set (Extended Data
Fig.~5) is the most conservative of the four and returns six descriptors recovered
by every near-optimal model.

This result is a precondition for the rest of the paper rather than a finding
alongside them. The route that is standard in the field returns, at a rate below
chance, precisely the descriptors that do not survive the boundary just measured.
No attribution method therefore licenses a statement about what will transfer. That
is why every boundary in this paper is crossed and measured rather than predicted
from the descriptors a model relies on within its own corpus.

\subsection*{The falsification test: an information-limited gap that a
representation does close}

The distinction organising this paper is worth nothing if it cannot fail. It would
fail if a gap could be closed both by buying target-side observations and by
changing the representation, because then the two diagnoses would not be
alternatives. The condition was fixed before the analysis: we needed a gap of the
second kind --- one that more data does not close --- and had to show that a
representation closes it. This section reports the case we found, at the length the
falsification test deserves. Without it, every failure in this paper could be
relabelled information-limited after the fact, and the framework would be
unfalsifiable.

On the Soundtracks corpus, which carries eight rating scales rather than two,
acoustic descriptors resolve tension ($\rho = 0.804$), tenderness ($0.754$),
valence ($0.735$), fear ($0.717$) and anger ($0.715$). Perceived happiness is the
single outlier at $0.474$ (Fig.~\ref{fig:tonality}a, open circles).

Musical happiness depends on mode and harmony, which the spectro-temporal
descriptor set does not represent. We therefore added 39 tonal descriptors: a
key-invariant chroma profile obtained by estimating the tonic with
Krumhansl--Kessler profiles \citep{krumhansl1982} and rotating, a mode score,
chroma flux and entropy, tonal centroids, an interval-consonance measure weighted
by a fixed a-priori consonance ordering, and sensory roughness computed from
spectral peaks after Plomp and Levelt \citep{plomp1965}.

The criterion was fixed before the test. Adding 39 descriptors will improve
performance somewhere; the question was whether the gain would concentrate on
happiness. It does: $+0.132$ for happiness against a mean of $+0.045$ for the other
seven axes, and the ordering of gains follows mode-dependence --- happiness,
valence and sadness lead; energy and anger, which depend on level, roughness and
rhythm, gain least (Fig.~\ref{fig:tonality}a). The deficit was representational
rather than dimensional: 39 tonal descriptors alone reach $\rho = 0.580$ for
happiness, above the 122 spectro-temporal descriptors at $0.474$
(Fig.~\ref{fig:tonality}b).

The mechanism check dissociates (Fig.~\ref{fig:tonality}c). Among all 161
descriptors, major-key strength and mode score rank 1st and 2nd for happiness but
3rd and 20th for tension, whereas interval consonance and sensory roughness rank
4th and 7th for tension but 117th and 76th for happiness. Mode carries perceived
happiness; consonance and roughness carry perceived tension. The ordering was
selected by the model, not imposed.

Because environmental recordings largely lack stable tonal structure, we asked
whether tonality is what fails to transfer across the domain boundary --- which
would give the barrier a mechanistic account rather than a descriptive one. It is
not. Adding tonal descriptors lowers music-to-environmental transfer by $0.021$
(18 paired comparisons, $p = 0.038$ uncorrected), in the predicted direction but at
a magnitude that accounts for roughly 4\% of the barrier itself (a fall from
$0.592$ to $0.054$). The reason is visible in the preceding paragraph: tonal
descriptors contribute almost nothing to \emph{arousal} specifically ($+0.006$
within the music domain, and $+0.012$ for the energy axis in
Fig.~\ref{fig:tonality}a), and arousal is the target on which the barrier was
measured. Tonality is thus one more descriptor family that does not transfer, not
the explanation for why none of them do. We therefore continue to report the
barrier as a phenomenon and make no mechanistic claim.

\subsection*{Boundary 2, cross-synthesis: direction survives the edit, calibration
does not}

A deployed system never plays the corpus. It plays audio that has been filtered,
compressed, layered or generated, so a synthesis boundary is crossed before any
listener hears anything --- and unlike the other three, this one is crossed by the
system's own design decisions rather than by circumstance. It is also the only
boundary on which we can intervene rather than observe, which is what makes it
informative about the boundary before it.

The preceding results are correlational. To ask whether the descriptors the models
rely on are also ones that \emph{move} a prediction when altered, we edited the
audio itself: 80 source excerpts $\times$ three interventions $\times$ four doses,
with the model held fixed and only the input changed
(Fig.~\ref{fig:intervention}).

Two of the three interventions passed their manipulation check --- that is, the
edit moved the descriptor it was designed to move, monotonically across dose
(Fig.~\ref{fig:intervention}b). Dose--response was then tested at the level of the
individual excerpt: a slope per excerpt across its four doses, and a Wilcoxon
signed-rank test of those 80 slopes against zero, with a sign-flip permutation
null. The modelled quantity throughout is the predicted probability of the
low-arousal class, so a positive slope means the edit made the excerpt read as
\emph{calmer}. Smoothing spectral flux, which lowers the rate of timbral change,
lowered predicted arousal in both domains (median slope $+0.015$ on that
probability, $p_{\text{perm}} = 1\times10^{-4}$ in the music domain; $+0.074$,
$1\times10^{-4}$ in the environmental domain), and did so consistently in sign
across 70\% and 71\% of excerpts. The directions agree across the domain boundary
that Fig.~\ref{fig:wall} shows predictions cannot cross, which is informative in
itself: \textbf{the barrier is one of calibration, not of sign.} A model fitted to
music does not know how loud is loud in an environmental recording, but it still
knows that smoother is calmer.

Injecting transients moved the probability the other way, as designed, and reached
significance ($p_{\text{perm}} = 1\times10^{-3}$ and $8\times10^{-4}$), but moved
only 50\% and 59\% of excerpts in a consistent direction --- at or barely above
chance. We therefore report it as a mean shift without a reliable per-excerpt
effect, and do not build on it. Envelope compression did not reach its intended
descriptor change and is reported as a failed manipulation rather than as a null
effect.

An earlier version of this analysis averaged the 80 excerpts within dose and
correlated the four resulting points, reporting rank correlations of $+0.632$ and
$+1.000$. That statistic cannot support the claim: with $n = 4$ the smallest
attainable two-sided $p$ for a Spearman correlation is $0.083$, so it could not
have reached significance whatever the data showed. The excerpt-level test above
replaces it.

The specificity diagnostic bounds what may be claimed
(Fig.~\ref{fig:intervention}c). No intervention moved a single descriptor in
isolation; each shifted a correlated family. The result therefore supports
causal statements at the level of a descriptor \emph{dimension} --- rate of
timbral change, onset density --- and not at the level of any individual
descriptor. It is also a statement about model behaviour: the listener response
to these edits was not measured here.

Of the three quantities we report at every boundary, this one supplies two. The
ceiling is the within-corpus performance of the model being perturbed, and the
survival is the sign agreement just described. The price is not defined here: there
is nothing to buy, because the deficit is not a missing quantity of data but a
missing calibration, and the calibration is supplied by the target-side labels
already priced at the previous boundary. What this boundary contributes instead is
a constraint on what those labels have to do. They do not have to teach the model
which acoustic direction is calming; it already knows. They have to tell it where
the scale sits.

\subsection*{Boundary 3, cross-response-channel: what listeners say and what their
bodies do}

The two boundaries reported so far are crossed by changing the sound. This one is
crossed by changing what counts as the response, while the sound and the listeners
stay fixed. It is the boundary an application crosses when it stops asking people
how they feel and starts reading a sensor instead, and it is the point at which the
budget-limited pattern of the first boundary stops holding.

\paragraph{Admission: what was excluded before the question was asked.}
Two corpora were removed before any result was computed, on criteria that had
nothing to do with what they might have shown. In the EEG corpus, the documented
mapping from stimulus trigger code to audio file does not reproduce, and an attempt
to recover it by matching emotion profiles did not beat its permutation null
($p = 0.19$); acoustics-to-physiology is therefore not testable there, and we do not
test it (Supplementary Section~1). In an electrodermal corpus of 94 participants,
no participant passed both pre-declared positive controls, so the pipeline was
withdrawn rather than reported (Supplementary Section~2). Neither exclusion is a
negative result. Both are the reason the negative results that follow can be read
at all: what remains is the subset in which a failure to find something means
something.

We analysed an EEG corpus of 31 listeners hearing 12\,s musical excerpts, recorded
with standardised event markers
\citep{daly2018data,gorgolewski2016,pernet2019}, processed with MNE-Python
\citep{gramfort2013}.

Four dataset-level positive controls were run before any substantive analysis
(Fig.~\ref{fig:channel}c). Relative occipital alpha exceeded frontal alpha in
31 of 31 listeners ($p = 1.9 \times 10^{-9}$), verifying channel identity and
spectral estimation. Averages time-locked to annotated blinks reached 164\,\uV{}
frontally, 3.98-fold the occipital amplitude and far above random windows
($p = 9.3 \times 10^{-10}$); because blinks are large and their times are annotated
independently, this verifies that event onsets index the correct samples. A
sound-onset response is present at 372--588\,ms, with peak $|t| = 8.78$ at 452\,ms
against a subject-wise sign-flip null with max-statistic correction
\citep{maris2007} ($p = 0.0001$); the pre-stimulus interval is not significant
($p = 0.54$), excluding a slow drift (Fig.~\ref{fig:channel}d). No classical N1
is present, while the stimuli have sharp onsets (median 25\,ms to half-amplitude),
so the absent fast component is attributed to playback-latency jitter; this limits
millisecond-scale analysis but not trial-wise spectral measures.

Against these controls, stimulus-level reliability was estimated for eight
self-report scales and 156 EEG measures on the same listeners, the same 1,240
trials and with the same estimator (Fig.~\ref{fig:channel}e). The EEG set spans
regional band power (26), magnitude-squared coherence between region pairs (50),
imaginary coherency (50) and electrode-pair asymmetry (30); coherence and asymmetry
are included because the original analysis of these data located its effect there
\citep{daly2014}. Coherence is indeed the strongest EEG family (maximum intraclass
correlation $0.092$, 95\% CI $[0.037, 0.142]$) and band power the weakest
($0.040$), against $0.221$ $[0.155, 0.281]$ for the best-agreeing self-report scale
on the same trials; the two intervals do not overlap.

That maximum is drawn from 156 measures, so it needs a null built the same way. We
permuted the stimulus labels, recomputed all 156 intraclass correlations and took
the largest, 400 times. Referred to that distribution the observed value carries
$p = 0.02$; referred to Benjamini--Hochberg across the same 156 it does not survive
(smallest $q = 0.16$) \citep{benjamini1995}. The two corrections disagree because
the value sits between them, not because either is wrong --- at rank one the
Benjamini--Hochberg threshold is the Bonferroni threshold, and the observation
falls just either side of it.

We therefore report a bound rather than an absence. The estimate is the maximum of
a selection and is biased upward for that reason, and $0.092$ is close to the
$0.10$ this design can resolve with 80\% power, so the interval is consistent with
values from near zero up to about a tenth of the response variance. What the data
do support, and what matters for the boundary, is the comparison: whatever
stimulus-specific structure the EEG carries is less than half of what the ratings
carry on the same trials, and the two intervals are disjoint.

No price could be established for this boundary, and the reason is worth stating
precisely, because it is weaker than a demonstration of absence. Across the
acoustic principal components, the ratings predicted from acoustics, and the
ratings themselves, no information source exceeds 31\% of the attainable ceiling on
the physiological target: the best of them is the subjective valence rating against
skin-conductance response rate, at $\rho = 0.117$ against a ceiling of $0.382$, or
$30.6\%$. Ratings predicted from acoustics reach $24.5\%$ and the leading acoustic
principal component $21.3\%$. Their confidence intervals overlap, so we do not
separate them, and the comparison that matters is against the control rather than
against each other: a random projection of the descriptor space reaches $20.6\%$ at
its 95th percentile, which places the two acoustic routes at the edge of what an
uninformative direction achieves and leaves only the human rating clearly above it.
When the best available predictor, the predictor derived from it, and the human
judgement it was derived from all stop within a third of the ceiling, the limit is
unlikely to lie in the predictor.

That is an argument from three routes failing alike, not from a saturation curve.
The other budget-limited boundaries were priced by buying target-side observations
and watching the recovery flatten; no equivalent curve exists here, because the
song-level electrodermal corpus does not support one. We therefore report that
every representation and every predictor we tested failed to recover the gap, and
we do not claim that no quantity of target-side data could. Distinguishing the two
requires a target-side learning curve on a physiological outcome, which is the
single most informative experiment this work does not contain.

Because the intended application is per-person rather than population-level, the
same question was asked within listeners, with the null distribution drawn from
each listener's own label permutations so that relationships differing in form
across people would not cancel. An individual-level positive control ---
predicting trial index, which must be predictable because impedance and vigilance
drift within a session --- passed for 29 of 31 listeners ($\rho = 0.573$ against a
null of $-0.222$). Of 248 listener $\times$ scale tests, 9 reached $p < 0.05$
against a chance expectation of $12.4$ (binomial $p = 0.88$), and none survived
correction (Extended Data Fig.~6).

Independent-component removal of ocular artefacts acts here as a sensitivity
manipulation check rather than as cleaning. It increased detection of the known
effect (individual-level control: 26 $\rightarrow$ 29 listeners passing,
$\rho = 0.401 \rightarrow 0.573$) while further reducing the effect of interest
(13 $\rightarrow$ 9 nominal detections). A pipeline made measurably more sensitive
did not reveal the association, which excludes insufficient signal-to-noise as an
explanation.

It does not, however, exclude insufficient observations, and the two must be kept
apart. The simulation reported below shows that with the roughly 30 rated trials
each listener contributed, a within-person association of $r = 0.40$ would have
been detected on 13\% of occasions, and one of $r = 0.20$ on 5\% --- that is, at
chance. \textbf{The individual-level analysis is therefore uninformative about
within-person associations below approximately $r = 0.5$, and we do not interpret
it as evidence of absence.} The between-listener result above does not share this
limitation, because it pools 31 listeners at the level of the stimulus; it is the
individual-level question, not the group-level one, that the design cannot address.

We also verified that the same conclusion is not an artefact of a single
electrodermal corpus that we had previously analysed \citep{kutt2022}: applying
individual-level positive controls to 94 of its participants, habituation of skin
conductance was in the expected direction for only 35\% of individuals, and no
participant passed both controls, so no usable subset exists (Supplementary
Information S2; recording standards after \citealp{boucsein2012}).

\subsection*{Boundary 4, cross-individual: for sale, at a price that depends on the
domain}

The last boundary is the one an application crosses when it stops serving people in
general and starts serving one person. Unlike the boundary before it, this one has
a price. Unlike the first, the price is not a single number.

To place the preceding limitation on a quantitative footing, synthetic within-person
associations of known strength were injected into data generated from the empirical
feature covariance of the EEG corpus (Ledoit--Wolf shrinkage \citep{ledoit2004})
and recovered with the identical pipeline, criterion and permutation test
(Fig.~\ref{fig:design}b). The requirement is steep: 314 observations per person for
$r = 0.40$ and 564 for $r = 0.30$, against the $\sim$30 this corpus supplies
(Table~\ref{tab:power}).

\begin{table}[htbp]
\centering
\small
\caption{Observations per person required for 80\% power, and the power actually
available at the $\sim$30 rated trials this corpus provides. Simulated rows use
injected associations of known strength; measured rows use the auditory evoked
response in an independent corpus of 75 listeners. The simulation was run at
500 replicates per cell with $n$ extended to 5{,}000, which resolves the two
weakest effects that a coarser grid could only bound at $> 1{,}000$.}
\label{tab:power}
\begin{tabular}{llcc}
\toprule
& Effect & Observations for 80\% & Detection at $n = 30$ \\
\midrule
\multirow{4}{*}{Simulated}
 & $r = 0.40$  & 314         & 0.10 \\
 & $r = 0.30$  & 564         & 0.07 \\
 & $r = 0.20$  & $1{,}364$   & 0.06 \\
 & $r = 0.10$  & $4{,}888$   & 0.04 \\
\midrule
\multirow{3}{*}{Measured}
 & Scalp EEG (N1)  & 57  & 0.53 \\
 & Pupil diameter  & 109 & 0.39 \\
 & Heart rate      & 233 & 0.23 \\
\bottomrule
\end{tabular}
\end{table}

Because that estimate rests on an assumed covariance structure, we measured the
same curve directly. An independent corpus of 75 listeners
\citep{ds003690,ribeiro2019erp,ribeiro2019cardiac} provides $\sim$240 auditory
events per person with
simultaneous scalp EEG, electrocardiogram and pupillometry, and carries an effect
that is certainly present: the evoked response to a tone. Establishing it at the
group level first is what makes the individual-level result interpretable --- at
$p = 0.0002$ for all three channels, an individual in whom it is not detected is
an individual in whom the measurement failed, not one in whom the effect is
absent. The cardiac component of that control, an anticipatory deceleration of
$-0.83$\,bpm following the cue, reproduces the effect those authors originally
reported in this cohort \citep{ribeiro2019cardiac}.

Detection was then measured per person as a function of how many of that person's
observations were used, with an identical sign-flip test at every point
(Fig.~\ref{fig:design}a). At the full 240 observations the evoked response is
recovered in 72 of 75 listeners from EEG, 62 of 73 from pupil diameter and 47 of
75 from heart rate (Benjamini--Hochberg corrected). At 30 observations --- what a
laboratory session supplies --- the same effect is recovered in 53\%, 39\% and
23\% of individuals respectively, while the false-positive rate stays at nominal
across the whole range (0.03--0.07). An effect plainly visible in the group
average ($-5.6\,\uV$ at the N1) is thus a coin toss within a single person at the
observation count these paradigms provide.

Three features of this measurement bear on how the simulated requirement should be
read. First, it validates the extrapolation itself: predicting detection at $n$
from an effect size estimated on a disjoint half of the same person's trials
matches the observed rate to within $0.03$ on average, so the procedure behind
Table~\ref{tab:power} is not an artefact of its distributional assumptions.
Second, the requirement is channel-specific, and in the direction that matters for
applied work --- heart rate, the signal a consumer device actually records, needs
roughly four times as many observations as scalp EEG (233 against 57). Third, the
requirement is not worse in older listeners; the evoked response is larger in the
older group, so fewer observations are needed (48 against 69 for EEG).

Comparable public corpora provide 40--60 observations per person and are in the
same regime. The estimates are lower bounds. The simulation assumes multivariate
normality, a linear association and within-person stationarity; each assumption
favours detection, and the last is the least tenable over long recordings, where
electrode drift and state change are larger. The measured curve is a lower bound
for a different reason: it uses a deliberately plain measurement --- a
fixed-window mean over one electrode cluster, with no independent component
analysis or spatial filtering --- so a dedicated pipeline would do better. It is,
however, representative of what a non-specialist pipeline achieves, which is the
situation an applied system is in.

One further result from that corpus is relevant to the choice of channel. In its
passive block, where the same tones are presented with no task and no response,
the cortical response is undiminished ($-5.7\,\uV$, $p = 0.0002$) but neither the
cardiac nor the pupillary response is detectable ($p = 0.30$ and $p = 0.99$). This
is not a power failure: had the effect been the size it is in the active blocks,
the group test would have detected it with probability $1.00$. Task relevance,
rather than the sound itself, appears to drive the peripheral channels here. The
scope of that observation is narrow --- 250\,ms tones in awake listeners, 30
passive trials each --- and it does not transfer to music heard over minutes. It
does mean that for a passively listening user monitored through heart rate alone,
the question of how many observations are needed may be preceded by the question
of whether there is a response to count.

\paragraph{The price is domain-dependent, and lowest in the domain nearest an
application.} The counts above are for detecting a within-person association at
all. The question an application actually faces is narrower: at what point does
fitting a listener their own slope beat giving them the population slope? That
break-even is reached when the between-person spread of the slope exceeds the error
in estimating it, and both quantities are measurable. We estimated them on three
corpora with a common estimator and, for each cell, a null distribution built by
parametric bootstrap from that cell's own design --- because a between-person
standard deviation estimated from few observations per person is biased upward, and
at 18 observations per person the estimator returns 0.11 when the true value is
zero (Table~\ref{tab:individual}).

The price is not one number. In urban soundscape ratings --- the setting closest to
an application that plays sound to change how a place feels --- the median cell
requires 67 observations per person, and two of eight cells have already crossed
break-even at the 42 observations the corpus supplies, both on the eventfulness
axis. In music ratings the median is 166. In song-level electrodermal responses to
music it is 318, and 11 of 15 cells cannot be distinguished from their own null at
all. Referred to a common 18 observations per person, the three settings clear
their nulls in 100\%, 50\% and 13\% of cells.

Four limits keep the two cells that cross break-even a candidate signal rather
than a validated result, and we would not act on them as they stand. The null for
each cell rests on 60 bootstrap replicates, so the Monte-Carlo error on the
threshold is of the same order as the margin by which the marginal cells clear it.
The correction subtracts the null in the variance metric, which is an empirical
device validated against simulation rather than a distributional result. No
multiplicity control is applied across the 33 cells, so with two crossings the
family-wise statement is weaker than the per-cell one. And the slopes are
estimated in two stages --- per person, then pooled --- where a cross-classified
hierarchical model would propagate the first stage's uncertainty into the second
instead of treating it as known.

The corpus is the fifth limit and the largest. ARAUS \citep{ooi2024araus} is
ratings of urban
soundscapes by awake listeners in a laboratory; it is nearer an application that
plays sound than any music corpus we have, and it is still not a sleep setting, a
long recording, or a listener whose state is drifting. By this paper's own
argument that is another boundary crossing, and an untested one. The defensible
reading is that per-person calibration first becomes worth testing in the
product-adjacent domain, at a scale existing corpora already reach --- not that it
has been shown to pay.

\begin{table}[htbp]
\centering
\small
\caption{When per-person calibration begins to pay. For each descriptor--outcome
cell, the between-person standard deviation of the slope was estimated by
restricted maximum likelihood and corrected against a null built by parametric
bootstrap from that cell's own design; $n^{*}$ is the number of observations per
person at which that spread exceeds the error in estimating it. Cells whose
estimate does not exceed their own null are counted separately, because a
between-person spread estimated from few observations per person is biased upward.}
\label{tab:individual}
\begin{tabular}{lccccc}
\toprule
Setting & Obs./person & Cells & Above own null & Median $\tau$ & Median $n^{*}$ \\
\midrule
Urban soundscapes, ratings & 42 & 8 & 8 & 0.133 & \textbf{67} \\
Music, ratings & 44 & 10 & 10 & 0.074 & 166 \\
Music, electrodermal & 18 & 15 & 4 & 0.000 & 318 \\
\bottomrule
\end{tabular}
\end{table}

Two steps of comparable size separate these: moving from soundscapes to music costs
about as much as moving from ratings to electrodermal response. Choosing the wrong
corpus is as expensive as choosing the wrong channel, which is not how the
literature on either is usually written. It also means our own earlier estimates,
made on music corpora, were pessimistic about the product-adjacent case by roughly
a factor of two.

\section*{Discussion}

\subsection*{What the results establish}

``The model does not generalise'' is not one diagnosis. It is two, and they call
for opposite remedies (Fig.~\ref{fig:boundaries}).

Taken in order, the four boundaries divide as follows. The first, cross-corpus,
contains both kinds inside a single experiment: swapping a corpus within a domain
costs a fifth of attainable performance and the loss is for sale --- a hundred
target-side labels return two-thirds of it --- while swapping across the domain
boundary costs four-fifths to all of it and four pretrained representations,
including one supervised on a corpus full of environmental sound, do not recover
it. The second, cross-synthesis, is not a loss of information but of calibration:
the sign of an acoustic edit's effect crosses the boundary that the predictions
themselves cannot. The third, cross-response-channel, is the one where no purchase
was found: no predictor we constructed exceeds 31\% of the attainable ceiling on
the physiological target, and the acoustic principal components, the ratings
predicted from them, and the ratings themselves all stop within a third of it. We
label it information-limited on that pattern, while noting that the label rests on
routes tried rather than on a saturating budget curve. The fourth,
cross-individual, is budget-limited with a price that varies by a factor of five
across settings, and in the setting closest to an application it has already been
paid.

The distinction is not a taxonomy imposed after the fact. It was stated as
falsifiable in advance: a boundary that could be closed both by more observations
and by a better representation would break it. We looked for that case and report
it at length --- perceived happiness, an information-limited gap that 39 tonal
descriptors do close --- because a framework in which every unrecoverable failure
can be relabelled after the fact is not a framework.

At the level of group-averaged ratings, acoustic descriptors are close to the limit
imposed by annotator agreement, and the informative variation lies in
generalisation. Two boundaries are sharp. The first is between sound domains, and
it is asymmetric: environmental sound does not transfer into music at all, while
music transfers outward in proportion to how much atmospheric material it contains.
The second is between measurement levels: the same listeners, on the same trials,
show substantial stimulus-specific agreement in what they report and, at most, a
component less than half that size in what their EEG does --- one that sits at the
edge of what correction for selection across measures will pass.

The second boundary requires care in statement, in three respects. It is not a claim
that the brain does not respond to sound: the onset response in these same
recordings is large and highly reliable. What is weak is the
\emph{differentiating} component --- a stimulus-specific pattern shared across
listeners. The common response is strong; the differential response is not.

Nor is it a claim that the differential component is zero. The strongest of the
156 EEG measures reaches an intraclass correlation of 0.09 (95\% CI 0.04--0.14)
against 0.22 (0.16--0.28) for the best-agreeing rating scale on the same stimuli.
Referred to a null built by permuting stimulus labels and taking the maximum across
all 156 measures --- the correction the selection requires --- that value carries
$p = 0.02$: at the boundary, neither a clean positive nor a demonstrated zero. It is
also the maximum of a selection, so the point estimate is biased upward, and it
falls near the 0.10 this design can resolve with 80\% power. The defensible
statement is a bound, not an absence: whatever stimulus-specific structure the EEG
carries is less than half of what the ratings carry, and the two intervals do not
overlap.

Nor do the two levels carry equal evidential weight. The between-listener result
pools 31 listeners at the level of the stimulus and is adequately powered. The
within-listener result is not: with roughly 30 rated trials per person, an
association of $r = 0.40$ would have been detected on 13\% of occasions. That
analysis therefore establishes only that no large within-person association is
obvious, and we draw no conclusion from it about associations of the size one would
actually expect. Distinguishing these two statuses matters, because they call for
different responses --- the first is a finding, the second is a design problem.

\subsection*{Why the negative is interpretable}

Physiological negatives are difficult to interpret because failure of the
measurement and absence of the effect produce the same number. Three features of
the present analysis separate them. Event-to-sample alignment was verified directly
through blink-locked averages, which is a bookkeeping check rather than a
physiological assumption. Positive controls were run at the level at which the
conclusion is drawn --- individually, not only in aggregate. And artefact removal
was used as a manipulation check: a step that measurably increased sensitivity to a
known effect did not increase the effect of interest. The third converts ``we could
not detect it'' into ``we made the instrument more sensitive and still could not
detect it'', at the cost of one additional analysis.

Two further failure modes are worth separating from these, because they are often
conflated. Measurement failure and low signal-to-noise are addressed by the
controls above. \textbf{Insufficient observations are not}, and no amount of
artefact removal substitutes for them. It was only on computing the power available
at 30 trials per person that the individual-level analysis in this study turned out
to be uninformative for any plausible effect size --- a limitation invisible in the
analysis itself, which returned a clean null with passing controls. A negative
result accompanied by positive controls can still be a negative result about
nothing at all.

We would therefore suggest that a design-sensitivity calculation belongs alongside
positive controls whenever a physiological null is reported, and that it be
computed on the empirical covariance of the actual measurements rather than from a
nominal effect size, since the two can differ by an order of magnitude in the
required $n$.

\subsection*{Which gaps are worth paying for}

Reading the four boundaries as one price list is the practical use of the
distinction. Two of them are for sale. Crossing to a new corpus within a domain
costs on the order of a hundred target-side labels for two-thirds of the gap, and
the first hundred are far cheaper per label than the second. Fitting a listener
their own slope costs 67 observations in the setting closest to an application, 166
in music ratings, and 318 in electrodermal response --- a range wide enough that
the choice of corpus is a budgeting decision, not a convenience.

Two are not for sale, and the reasons differ. Crossing between sound domains is not
recovered by any of four pretrained representations, including the one that ought
to cross; what target labels buy there is the calibration the synthesis experiment
isolates, not the mapping, which already transfers in sign. Crossing to a
physiological response is the one boundary where the ceiling itself is low: the
acoustic principal components, the ratings predicted from them, and the ratings
themselves all stop within a third of it, so the limit is unlikely to lie on the
predictor and is not obviously moved by improving it.

A single rule follows. Before spending on more data or a larger model, establish
which kind of failure is at hand --- and the cheapest way to establish it is to
check whether a different information source, rather than more of the same one,
moves the number. Where it does not, the budget is better spent on a different
measurement than on a larger one.

The scope is specific: approximately 30 observations per person, 12\,s excerpts,
single exposure, scalp EEG, cross-sectional design. Each element is a property of
the design rather than of the phenomenon, and the simulation quantifies the first
of them directly --- 314 observations per person for $r = 0.40$, 564 for
$r = 0.30$, 1,364 for $r = 0.20$ and 4,888 for $r = 0.10$.

These numbers are not unreachable in absolute terms. They are unreachable within a
session \emph{that also varies its stimuli}. A recording that samples a listener
once per minute over an eight-hour night yields on the order of 480 observations,
so the $r = 0.30$ requirement corresponds to a little over one night and the
$r = 0.40$ requirement to less than one, while $r = 0.20$ would take about three
nights and $r = 0.10$ about ten. The gap between what the question
needs and what the paradigm supplies is thus about a factor of twenty, and it is a
gap in \emph{duration and repetition} rather than in participants: recruiting more
people does not help, because the quantity in short supply is observations within
a person.

The qualification matters, because observation count and stimulus richness trade
against each other inside a fixed session. The corpus used above for the measured
curve reaches 240 events per person in about 45 minutes --- but it does so with
four pure tones, which is to say with no acoustic variation to model. Rating
corpora make the opposite trade: hundreds of distinct excerpts, a few tens of them
per listener. What the question requires is both at once, and it is the
combination, not either quantity alone, that no session-length design supplies.

The relevant point is that this design is not incidental to the corpora we used but
structural to the paradigm that produced them. A single laboratory session is
bounded by participant fatigue and electrode drying; a stimulus set broad enough to
cover an affective space forces a trade-off between observations per stimulus and
stimuli per participant --- in the present corpus, 307 distinct stimuli with a
median of two listeners each and a median pairwise overlap of six excerpts; and a
cross-sectional design cannot separate state from trait because it never observes
the same person twice.

The data required to answer the question are therefore within-person,
longitudinal, varied in stimulus, and of an order of magnitude in observations per
person that no single session provides. We were unable to locate such a
combination in public form: a systematic search of OpenNeuro (1,833 datasets) and
PhysioNet (426) returned no corpus pairing repeated controlled auditory
stimulation with a state readout at the required density, and the standard
affective-physiology corpora distributed outside those repositories --- DEAP,
AMIGOS, MAHNOB-HCI, DREAMER, ASCERTAIN, CASE --- supply between 8 and 40 stimuli
per participant, one to two orders of magnitude below the measured requirement for
a peripheral channel. The claim is bounded by that search rather than absolute.
It is also a statement about what would settle the question, not a prediction
about how it would be settled: an association of the required size may or may not
be present once the observations are available.

\subsection*{Limitations}

Perceived arousal in a rating corpus is not sleep onset, sleep continuity or any
clinical endpoint; the affective annotations used here bear on the former only.
Of the three counterfactual audio manipulations (Fig.~\ref{fig:intervention}), one
--- spectral smoothing --- shifts model outputs in the predicted direction with a
per-excerpt effect that is both significant and consistent in sign; the other two
are a failed manipulation and a mean shift without per-excerpt consistency. Even
for the first, the claim is at the level of a descriptor dimension, has not been
verified in listeners, and the specificity of the manipulation does not support
claims about individual descriptors. Cardiac and pupillary signals
enter this work only through the design-requirement measurement, where the
stimulus is a pure tone; no analysis here relates a naturalistic soundscape to a
peripheral physiological measure, and respiratory and actigraphic signals are not
covered at all. Two properties of the tone corpus limit how far its numbers
travel: the listeners were awake and performing a task, and the events were
250\,ms tones rather than minutes of music. Finally, the acoustic-to-physiological
link
could not be examined in the EEG corpus at all: the mapping from stimulus trigger
codes to audio files given in the dataset documentation does not reproduce, and an
attempt to recover it from emotion-profile matching did not exceed its permutation
null ($p = 0.19$). The listeners' own ratings and their EEG remain analysable
because the trigger codes are internally consistent, but the audio itself could not
be linked. Two further discrepancies in that dataset's event annotation are
reported in Supplementary Information S1.

\section*{Methods}

\subsection*{Corpora}

Four rating corpora were used: DEAM (music, $n = 1{,}233$) \citep{aljanaki2017},
PMEmo (music, $n = 717$) \citep{zhang2018pmemo}, Soundtracks Set 1 (music,
$n = 360$) \citep{eerola2011} and Emo-Soundscapes (environmental, $n = 1{,}213$)
\citep{fan2017emo}. ESC-50 \citep{piczak2015} served as a second environmental
corpus for the category control. The prediction target throughout is
group-averaged perceived arousal, rescaled to $-1$--$1$; the circumplex framing
follows \citet{russell1980} and \citet{eerola2013}. Soundtracks additionally
carries seven further scales (valence, tension, anger, fear, happiness, sadness,
tenderness) on nine-point Likert scales, linearly rescaled to the same range --- a
monotone transformation that leaves all reported rank statistics unchanged.

Soundtracks was included specifically to test whether within-music transfer
reflects shared acoustic structure or a shared annotation tradition: it originates
from a different laboratory and decade, uses a categorical rather than continuous
rating protocol, and was assembled as a balanced design of twelve target emotions
$\times$ 30 excerpts.

\subsection*{Audio preparation}

All audio was resampled to 22{,}050\,Hz, centre-cropped to at most 30\,s, and
loudness-normalised to $-23$\,LUFS (ITU-R BS.1770) before any descriptor was
computed. Normalisation is not cosmetic: without it, corpus-specific mastering
levels act as an identifier that inflates apparent cross-corpus transfer (Extended
Data Fig.~1). For the duration control, music excerpts were additionally cropped to
a 6\,s centred window to match the environmental corpus.

\subsection*{Descriptors}

\paragraph{Spectro-temporal (122).}
Mel-frequency cepstral coefficients, spectral centroid, bandwidth, contrast,
flatness and roll-off, zero-crossing rate, root-mean-square level, onset strength
and onset rate, spectral slope, and band-energy ratios. Each frame-level series is
summarised by mean, standard deviation, 10th, 50th and 90th percentiles, and the
mean absolute first difference, the last of which indexes rate of change rather
than level. Computed with librosa \citep{mcfee2015}.

\paragraph{Tonal (39).}
Harmonic--percussive separation ratio; a key-invariant chroma profile obtained by
estimating the tonic through correlation of the mean constant-Q chroma vector with
Krumhansl--Kessler major and minor profiles \citep{krumhansl1982} and rotating the
vector so that the tonic occupies index 0; major-key strength, minor-key strength,
their difference (mode score) and key clarity; chroma flux and chroma entropy; six
tonal-centroid dimensions with standard deviations; an interval-consonance measure
formed by weighting each pitch-class pair by the consonance of its interval;
sensory roughness computed over the strongest spectral peaks per frame after Plomp
and Levelt \citep{plomp1965}; and fundamental-frequency median, interquartile range
and voiced fraction.

Rotation to the tonic is essential rather than convenient. An unrotated chroma
vector encodes which key a piece is in, which is arbitrary with respect to affect
--- transposing a piece does not change its emotional character --- so an unrotated
representation would let a model learn the key distribution of a corpus. The
interval-consonance weights are fixed a priori from the standard ordering of
sensory consonance (unison $>$ perfect fifth $>$ perfect fourth $>$ major third and
sixth $>$ minor third and sixth $>$ major second and minor seventh $>$ minor second
and major seventh $>$ tritone) and are not fitted, so they contribute no degrees of
freedom.

\paragraph{Pretrained embeddings (512).}
Audio embeddings from a contrastive language--audio model \citep{wu2023clap},
used only for the representation comparison (Extended Data Fig.~7). One corpus
derives from the public sound repository that appears in the model's pretraining
data, so the comparison is reported separately for it and for two corpora with no
such overlap.

\subsection*{Models and evaluation}

Six regressors --- ElasticNet, support vector regression, $k$-nearest neighbours,
random forest, gradient boosting and XGBoost \citep{chen2016xgboost} --- were
evaluated by five-fold GroupKFold cross-validation with groups defined by source
recording, so that no excerpt from a recording appears in both training and test
folds. Reported values are Spearman rank correlations between predicted and
observed values; cross-corpus transfer fits on the whole of one corpus and
evaluates on the whole of another. Where a single value is reported for a cell, it
is the best of the six algorithms; where a comparison between conditions is made,
all six are retained as paired observations and the comparison is tested by
Wilcoxon signed-rank.

\paragraph{Annotation reliability ceiling.}
Per-rater judgements were split at random into two halves, stimulus means computed
within each, correlated, and corrected by the Spearman--Brown formula; the reported
value is the mean over 200 random splits. Model performance is expressed as a
fraction of this ceiling.

\subsection*{EEG acquisition and preprocessing}

Data are OpenNeuro ds002721 \citep{daly2018data}: 31 listeners, 19 scalp channels
at 1\,kHz referenced to FCz, six runs per listener of which four contained ten
12\,s music excerpts each. Event annotation follows BIDS and its EEG extension
\citep{gorgolewski2016,pernet2019}, which defines event onsets relative to the
first recorded sample --- the property that makes the alignment check below
possible. Processing used MNE-Python \citep{gramfort2013}: 50\,Hz notch, 1--45\,Hz
bandpass, resampling to 250\,Hz. Spectral and asymmetry measures were computed
after re-referencing to the average of the 19 channels; coherence was computed on
the recording reference, because average referencing introduces a common component
across channels that inflates coherence.

\paragraph{Trial-wise measures (156).}
Relative band power (delta, theta, alpha, beta, gamma) in five regions, expressed
as the change from a 2\,s pre-onset baseline to a 1--14\,s response window (26
measures including frontal alpha asymmetry); magnitude-squared coherence between
all ten region pairs in five bands (50); the imaginary part of coherency for the
same pairs and bands (50), which is insensitive to the zero-lag correlations
produced by volume conduction; and log-ratio asymmetry for six homologous electrode
pairs in five bands (30). Coherence and asymmetry families are included because the
original analysis of these recordings located its effect there \citep{daly2014};
restricting the measure set to band power would test the conclusion on a
representation that excludes the effect in question.

\paragraph{Artefact handling.}
For the main analyses no epoch rejection was applied, so that the stimulus and null
arms receive identical treatment. In the sensitivity analysis,
independent-component analysis (15 components) was fitted per run and components
correlated with the frontopolar channels, used as ocular proxies, were removed
(median one component per run). All other parameters were held fixed, so that any
change is attributable to artefact removal alone.

\subsection*{Controls preceding physiological analysis}

Four checks were specified and run before any substantive analysis, with the
decision rule that failure of any one would halt interpretation.

\begin{enumerate}
\item \textbf{Event bookkeeping.} Each music run must contain exactly ten
music-onset markers, pair bijectively with ten stimulus codes, and carry exactly
eighty response events.
\item \textbf{Channel identity and spectral estimation.} Relative alpha power must
be greater at occipital than at frontal sites during rest.
\item \textbf{Event-to-sample alignment.} Averages time-locked to independently
annotated blink events must show a large frontal deflection relative to occipital
sites and to random windows. This is a bookkeeping check: it depends only on blinks
being large and their times being recorded, not on any physiological hypothesis.
\item \textbf{Stimulus-locked response.} An evoked response to sound onset must
exceed a subject-wise sign-flip null with max-statistic correction across the epoch
\citep{maris2007}.
\end{enumerate}

Control 4 initially failed under a criterion that compared evoked global field
power in a 0--300\,ms window against temporally shifted nulls. Two problems were
identified. The window was placed for a classical N1 while the response in these
data peaks at 452\,ms; and any temporally shifted null moves the analysis window
into the response-collection period, which has a different noise environment,
because stimulus onsets fall in the quietest part of the trial structure. The
sign-flip null avoids both by not moving in time. We report this because the
mis-specified criterion would, if followed, have led us to discard a valid dataset.

\paragraph{Individual-level controls.}
Because the substantive question is per-person, the same logic was applied within
listeners: each listener's EEG must predict trial index, which is necessarily
predictable given impedance and vigilance drift within a session. Listeners failing
this check would have their negative results treated as uninterpretable.

\subsection*{Reliability and decoding}

\paragraph{Stimulus-level reliability}
was estimated in two ways that agree. A one-way random-effects intraclass
correlation was computed over all 1,240 trials after within-listener z-scoring,
which is the appropriate estimator for the unbalanced design (307 distinct stimuli,
median two listeners each). Split-half reliability across listeners with
Spearman--Brown correction was computed on the subset of stimuli heard by at least
eight listeners ($n = 44$). Significance used within-listener permutation of
stimulus labels.

\paragraph{Per-listener decoding}
used five-fold cross-validation with standardisation and principal-component
reduction to eight components fitted within each training fold, followed by ridge
regression with cross-validated penalty. The evaluation metric is the rank
correlation between out-of-fold prediction and observation. This metric is biased
downwards under cross-validation at small $n$ --- training-fold means are displaced
from held-out values --- so it is never compared against zero but always against a
null formed by permuting that listener's own labels through the identical pipeline
(200 permutations). Leave-one-out cross-validation was tried first and abandoned:
it produced observed and null values both near $-0.5$, rendering the comparison
uninformative.

Multiplicity across measures was controlled by the Benjamini--Hochberg procedure
\citep{benjamini1995}.

\subsection*{Design simulation}

Feature covariance was estimated from the artefact-corrected EEG measures by
Ledoit--Wolf shrinkage \citep{ledoit2004}. For each combination of true association
strength ($r \in \{0.10, 0.20, 0.30, 0.40\}$) and observations per person
($n \in \{30, 60, 120, 250, 500, 1000, 2000, 5000\}$), 500 datasets were generated
from that covariance with a sparse linear association injected and scaled to the
target strength, and each was analysed with the identical pipeline, criterion and
permutation test used on the real data (100 permutations per dataset). Reported
power is the proportion of datasets in which the association was detected at
$p < 0.05$. An earlier run at 60 datasets per cell and $n \le 1{,}000$ could only
bound the two weakest effects at $> 1{,}000$; extending the grid resolves them at
1,364 and 4,888. The two strongest requirements shifted by less than the
Monte-Carlo error of the coarser run (271 to 314 and 650 to 564).

The simulation is optimistic by construction. Multivariate-normal generation
understates the heavy tails of real measurements; the injected association is
linear; and within-person stationarity is assumed, which is least tenable over the
long recordings the result is used to discuss. Required sample sizes are therefore
lower bounds.

\subsection*{Measured design requirement}

The simulated requirement was checked against a direct measurement in an
independent corpus of 75 adults (36 aged 19--30, 39 aged 50--70) recorded with
64-channel EEG, bipolar electrocardiogram and binocular pupillometry at 500\,Hz
during cued reaction-time tasks \citep{ds003690,ribeiro2019erp,ribeiro2019cardiac}.
Auditory cues were 250\,ms pure tones at approximately 67\,dB(A). Each participant
contributed 240 cues across two active tasks and 30 in a passive block. Event
timing was verified against the sample indices supplied in the BIDS event files.

Three response measures were fixed in advance, one per channel: the mean
fronto-central amplitude over 80--130\,ms after cue onset (nine electrodes,
mastoid reference, 0.1--30\,Hz, epochs exceeding $\pm 150\,\uV$ rejected); the mean
change in instantaneous heart rate over 1--5\,s relative to a 3\,s pre-cue baseline
(R peaks detected on a 5--35\,Hz band); and the mean change in pupil diameter over
0.5--2.5\,s relative to a 1\,s baseline, with blink samples interpolated and epochs
exceeding 30\% interpolation rejected. A pre-declared secondary EEG measure, the
difference between 170--230\,ms and 80--130\,ms, is reported alongside the primary
one; inspection of the pre-declared grand-average waveform showed the later window
to fall on the contingent negative variation rather than on a P2, as is expected in
a cued task, so the single-window measure was adopted as primary. Both are
fixed-window means, for which the sign-flip null below is exact; peak-to-peak
measures are not, being positive under the null by construction.

Each real cue was paired with a randomly placed anchor in the same recording, and
all analyses operate on the difference. Random anchors rather than a fixed temporal
shift were used because the inter-cue interval leaves no quiet window of the
required length; the difference also removes a monotone within-run decline in heart
rate that the anchor arm revealed. Under the null of no event-locked response the
two arms are exchangeable, so sign-flipping the paired differences is exact.
Detection was assessed per participant with 5,000 sign flips at $\alpha = 0.05$ and
Benjamini--Hochberg correction across participants. The trial-count curve
subsamples $n$ of a participant's pairs, repeats the identical test 40 times per
value of $n$, and reports the mean detection rate; the accompanying false-positive
rate applies the same procedure to sign-randomised differences, which preserves the
noise but removes the effect. Extrapolation was validated out of sample by
estimating each participant's effect size on a random half of their trials and
predicting detection on the other half.

Heart rate and pupil diameter were analysed separately by task, a stratification
fixed on design grounds: the active-task response window contains the imperative
tone and the button press, the passive block contains neither, and the two differ
eightfold in trials per person.

\subsection*{Supplementary electrodermal corpus}

An electrodermal corpus \citep{kutt2022} was analysed and its results withdrawn.
Processing used NeuroKit2 \citep{makowski2021} following standard recommendations
\citep{boucsein2012}. Group-level positive controls failed --- habituation of skin
conductance across trials was significant in the direction opposite to the
established effect, and post-stimulus windows did not exceed random windows. The
possibility that a usable subset existed, with group means displaced by equipment
failures in the remainder, was tested by applying both controls per participant: of
94 analysable participants none passed both, and habituation was in the expected
direction for only 35\%, so no subset exists. Details are in Supplementary
Information S2.

\subsection*{Reproducibility}

Each analysis writes a run record --- 172 of them --- carrying the script, the
random seed, the parameters and the resulting metrics, and source data for every
figure panel are provided as machine-readable tables. What that record does and
does not establish should be stated plainly, because the two are often conflated.

It fixes the seed, the parameters and the numbers, so a reader can see exactly
what was computed and compare it with the tables. It does not currently pin the
code: all 172 records were written from a working tree with uncommitted changes,
and the repository head is older than most of the analyses, so the recorded commit
identifies a starting point rather than the state that produced the number. Input
hashes are present for 38 of the 172 and absent for the rest; output hashes are
not recorded at all. There is no automated test suite. The figures are the
exception --- each figure script asserts the values the text quotes from it, so a
figure that drifts from the prose breaks the build --- but that discipline covers
the figures only.

The practical consequence is that the analyses are reproducible from the scripts
and the public corpora, at the parameters recorded, and are not yet reproducible
bit-for-bit from a pinned commit. Bringing the record up to the stronger claim
requires committing the tree, re-running under a clean checkout and recording
input and output hashes throughout; that work is not done, and we would rather say
so than let the weaker guarantee be read as the stronger one.

Reference metadata were verified individually by DOI lookup with reverse title
checking; entries that could not be confirmed were excluded rather than cited.

% =====================================================================
\section*{Data availability}

All analyses use publicly available data. Accession identifiers, access dates and
licence terms are listed below; each corpus is additionally cited in the text as
requested by its distributors.

\begin{itemize}
\item \textbf{ds002721} --- OpenNeuro accession \texttt{ds002721}, retrieved
2026-07-27, \url{https://openneuro.org/datasets/ds002721}. Note that the dataset's
\texttt{dataset\_description.json} records CC0 while its README records
CC~BY~4.0; we have followed the more restrictive terms and attribute the source
\citep{daly2018data}.
\item \textbf{Soundtracks Set 1} --- Open Science Framework accession
\texttt{p6vkg}, retrieved 2026-07-27, \url{https://osf.io/p6vkg/}. The deposit is
released under CC~BY~4.0 \citep{eerola2011}; the excerpts are nonetheless taken
from commercially released film scores, so the deposit's own licence does not
convey rights in the underlying recordings, and we treat the audio as usable for
research only.
\item \textbf{DEAM} --- distributed by the University of Geneva; audio licences
vary per track \citep{aljanaki2017}.
\item \textbf{PMEmo} --- annotations and precomputed features are redistributable;
the audio consists of commercial excerpts and is not \citep{zhang2018pmemo}.
\item \textbf{Emo-Soundscapes} --- assembled from Creative Commons material on
Freesound; per-item licences are given in the dataset metadata
\citep{fan2017emo}.
\item \textbf{ESC-50} --- CC~BY-NC~3.0 \citep{piczak2015}.
% ARAUS carries the cross-individual result and is CC BY-NC 4.0, whose BY term
% makes this entry an obligation rather than a courtesy. It was missing from
% this list while the paper reported n* = 67 from it.
\item \textbf{ARAUS} --- distributed by Nanyang Technological University via
DR-NTU, retrieved 2026-08-03. CC~BY-NC~4.0 \citep{ooi2024araus}. Used for the
cross-individual analysis only; the non-commercial term means the material
itself cannot be reused in an application, whereas the calibration requirement
estimated from it can.
\item \textbf{BIRAFFE2} --- CC~BY~4.0 \citep{kutt2022}; used only for the
methodological case reported in Supplementary Information S2.
\item \textbf{ds003690} --- OpenNeuro accession \texttt{ds003690}, retrieved
2026-07-30, \url{https://openneuro.org/datasets/ds003690}. CC0
\citep{ds003690}. The distributors request that the EEG and pupil data be
attributed to \citet{ribeiro2019erp} and the electrocardiogram to
\citet{ribeiro2019cardiac}; both are cited accordingly.
\end{itemize}

No new data were generated. Derived tables underlying every figure panel are
provided as Source Data. Because these are small tabular files rather than a
primary dataset, they accompany the manuscript rather than being deposited in a
community repository.

\section*{Code availability}

Analysis code, run records and figure-generation scripts are available at
\url{https://github.com/AuraJZ/affective-audio-boundaries}. Each reported value
has a run record identifying the script, the random seed and the parameters used;
the scope and the current limits of that record --- uncommitted working trees,
incomplete input hashes, absent output hashes --- are set out under
Reproducibility in the Methods, and this statement should be read against it
rather than as a stronger guarantee.

\section*{Ethics}

This work is a secondary analysis of de-identified data released for public reuse
by the original investigators, who obtained informed consent and ethical approval
as described in their respective dataset publications. No new data were collected
from human participants, and no attempt was made to re-identify any individual.

\section*{Competing interests}

The authors are engaged in the development of a commercial sleep-audio product.

\section*{Author contributions}

\textbf{Jingyi Zhang:} conceptualisation; methodology; software; validation;
formal analysis; investigation; data curation; visualisation; writing --- original
draft; writing --- review and editing.
\textbf{Xiaotong Yao:} conceptualisation; resources; validation; writing ---
review and editing; project administration.

\section*{Acknowledgements}

We thank the investigators who released the datasets analysed here. This work would
not have been possible without their decision to make the recordings, annotations
and event files public.

\bibliography{../references/references}

% =====================================================================
% FIGURES
%
% No \section*{Figures} heading and no \clearpage, deliberately. Every figure
% below is a [p] float, and a [p] float never shares a page with running text,
% so a heading here is guaranteed a page of its own with nothing else on it.
% Nor can the first figure be moved up to join it: at \textwidth Fig. 1 is
% 171 mm tall and its legend runs about 126 mm, against a 247 mm text block.
% The floats therefore begin straight after the references, each self-labelled.
%
% Figure order follows the argument: the ceiling, then each boundary in the
% order an application meets it, then the synthesis. Two figures are new to this
% version -- the price of the first boundary (Fig. 2) and the four-boundary
% summary (Fig. 8) -- and the rest are renumbered accordingly. The legend macro
% names track POSITION, not content, so they must be re-checked whenever the
% order changes; see reports/FIGURE_LEGENDS.md, which is the source of truth.

\msfigure{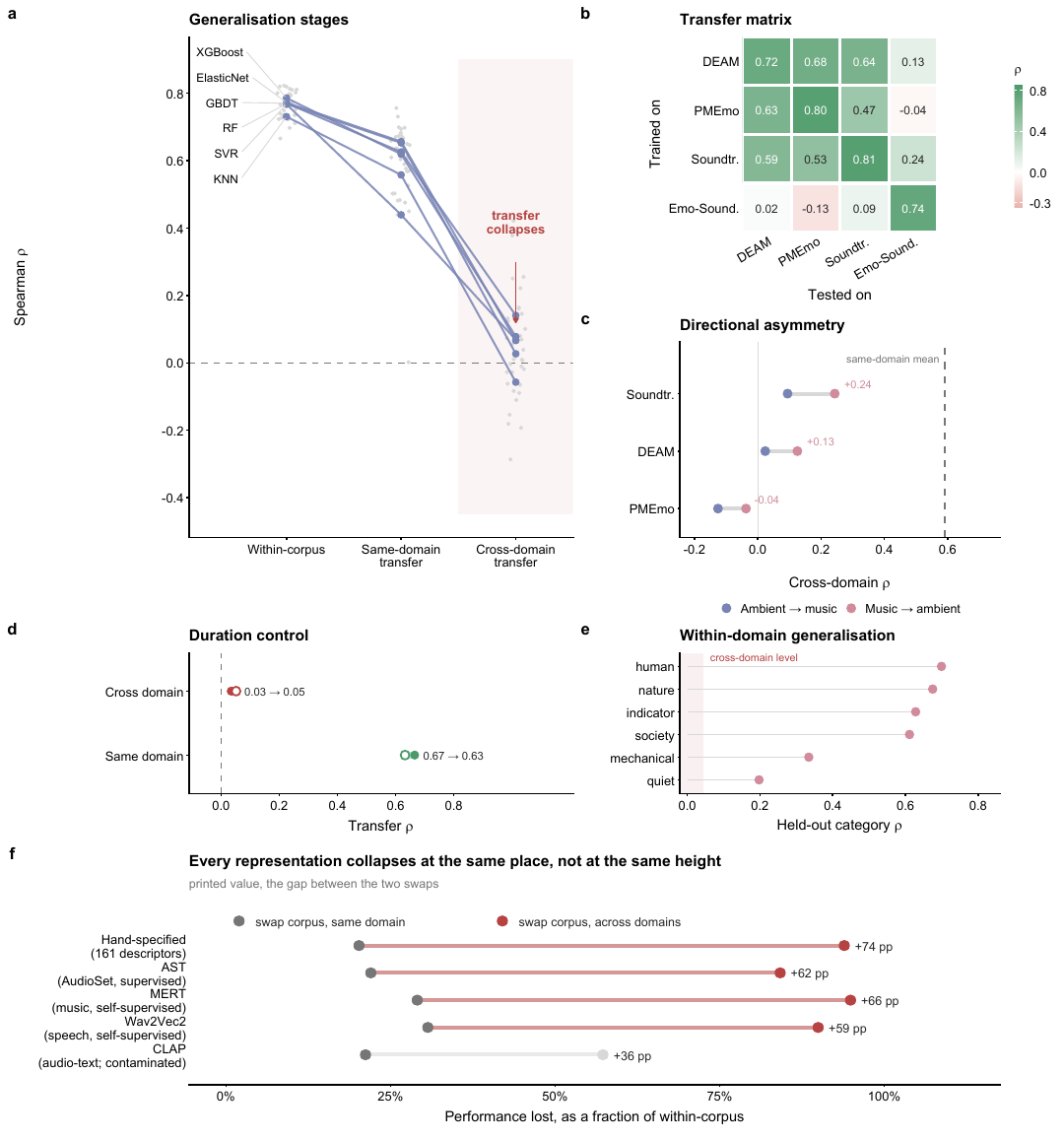}{fig:wall}{\legendfigone}

\msfigure{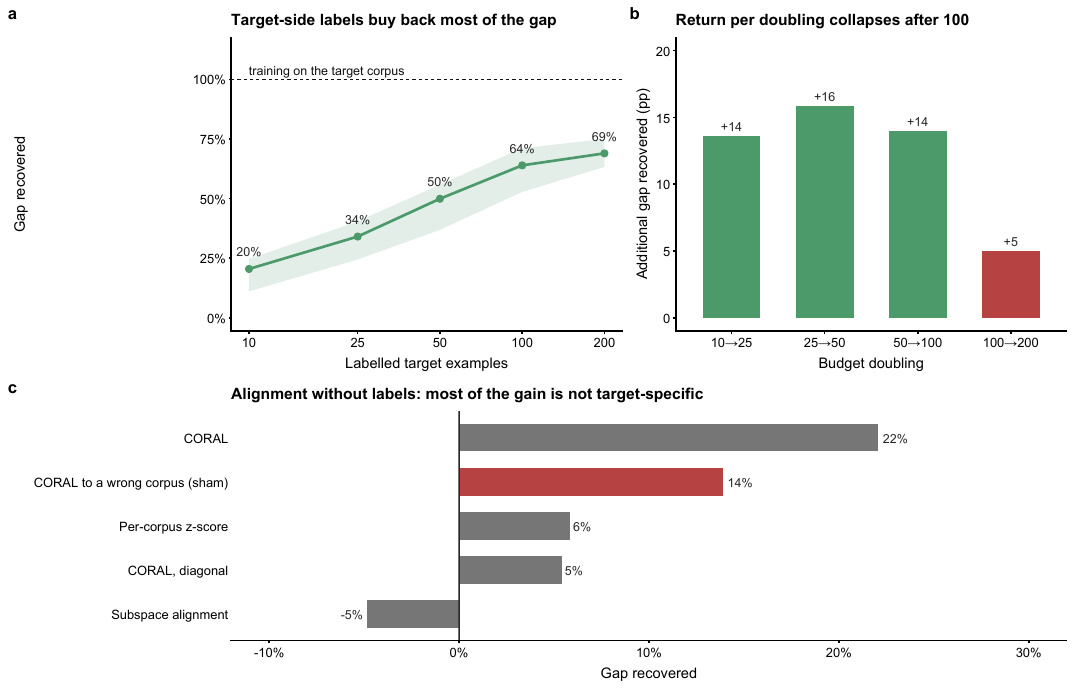}{fig:price}{\legendfigtwo}

\msfigure{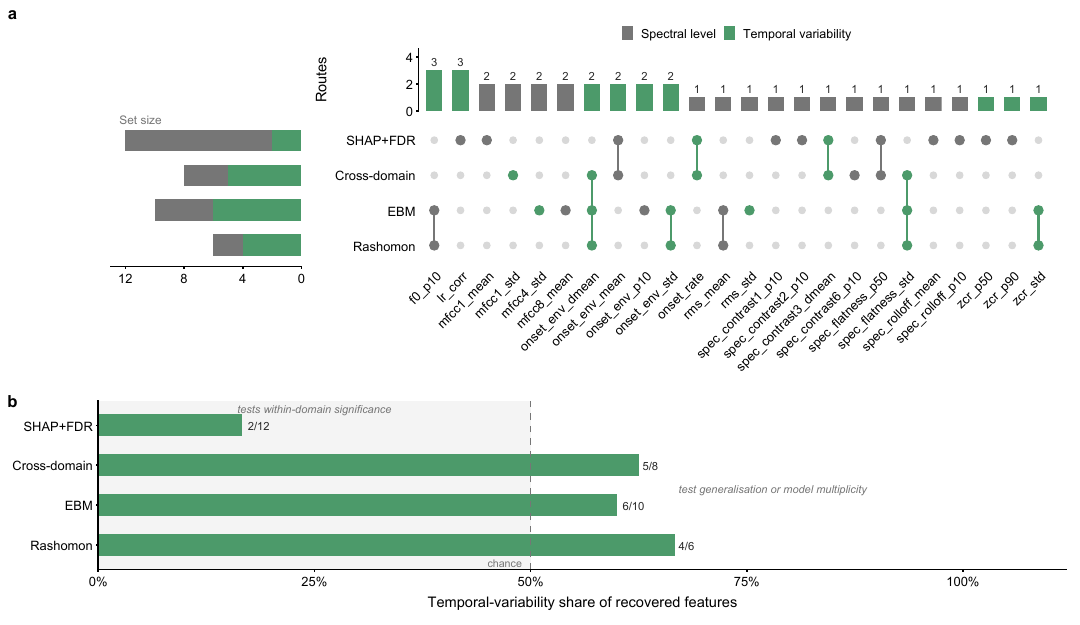}{fig:routes}{\legendfigthree}

\msfigure{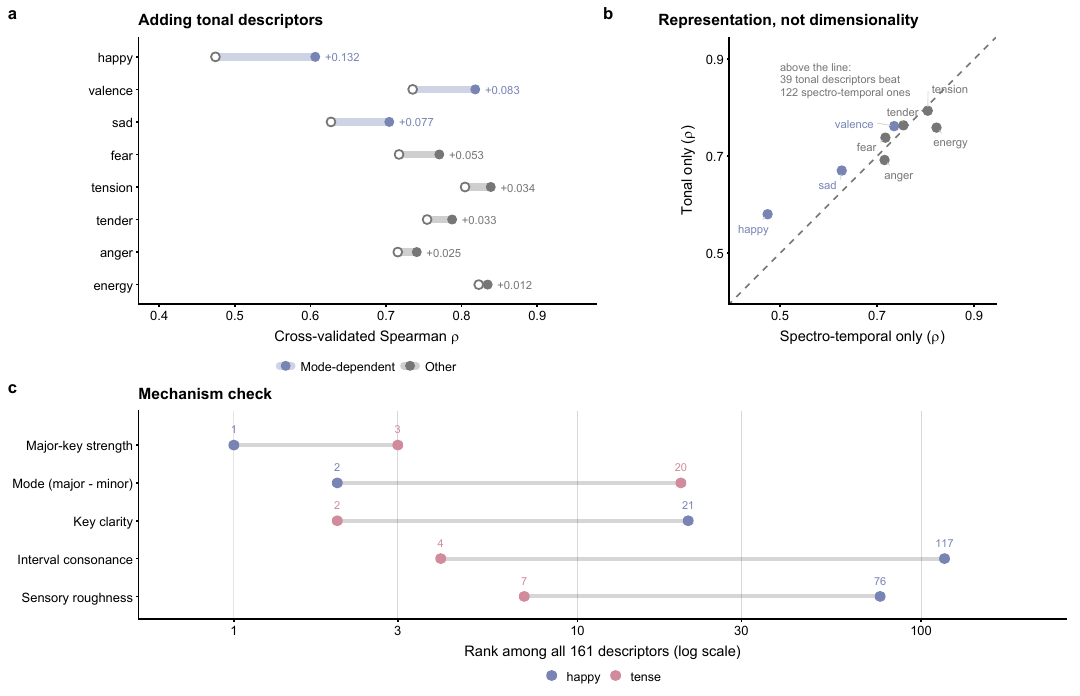}{fig:tonality}{\legendfigfour}

\msfigure{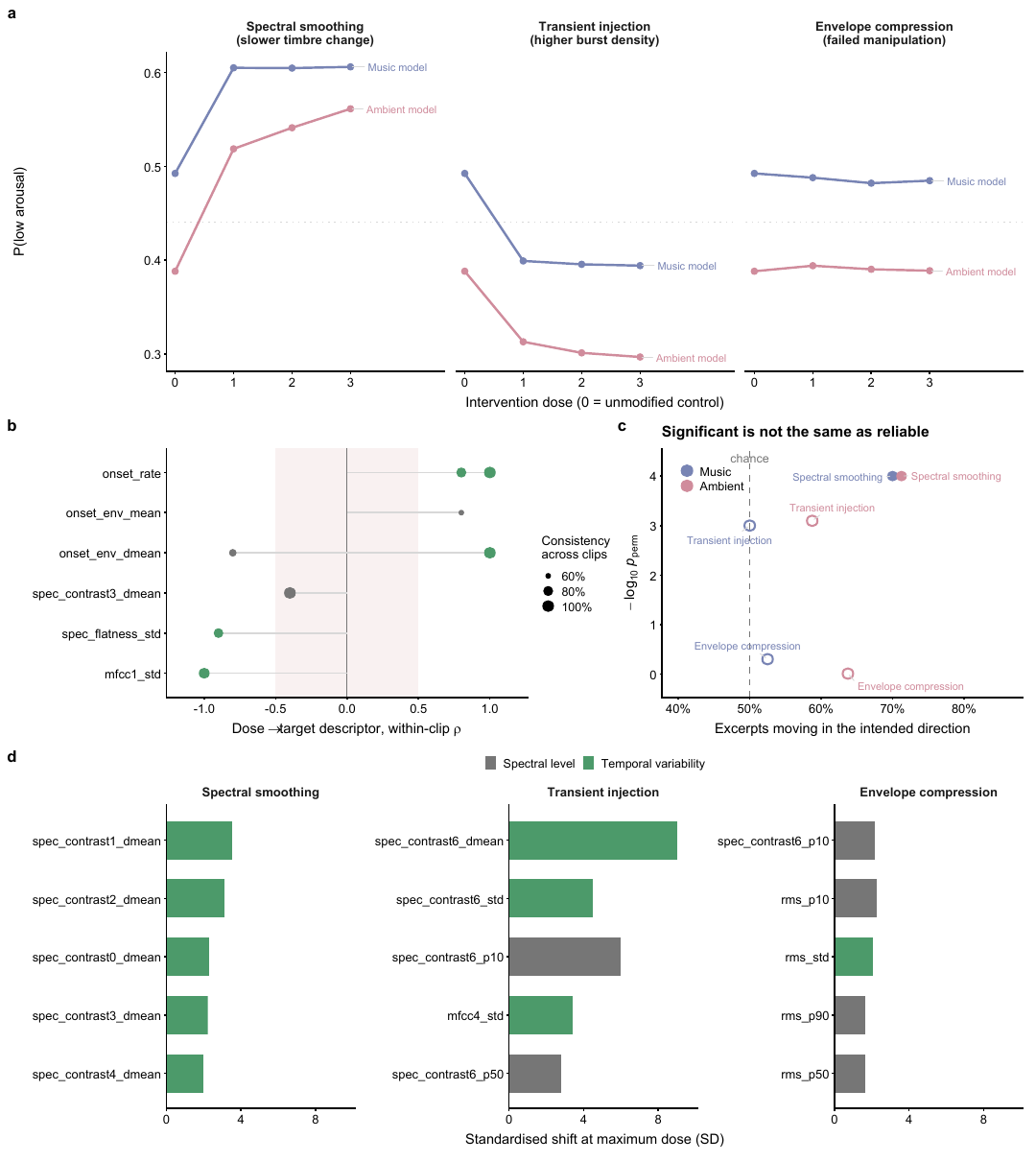}{fig:intervention}{\legendfigfive}

\msfigure{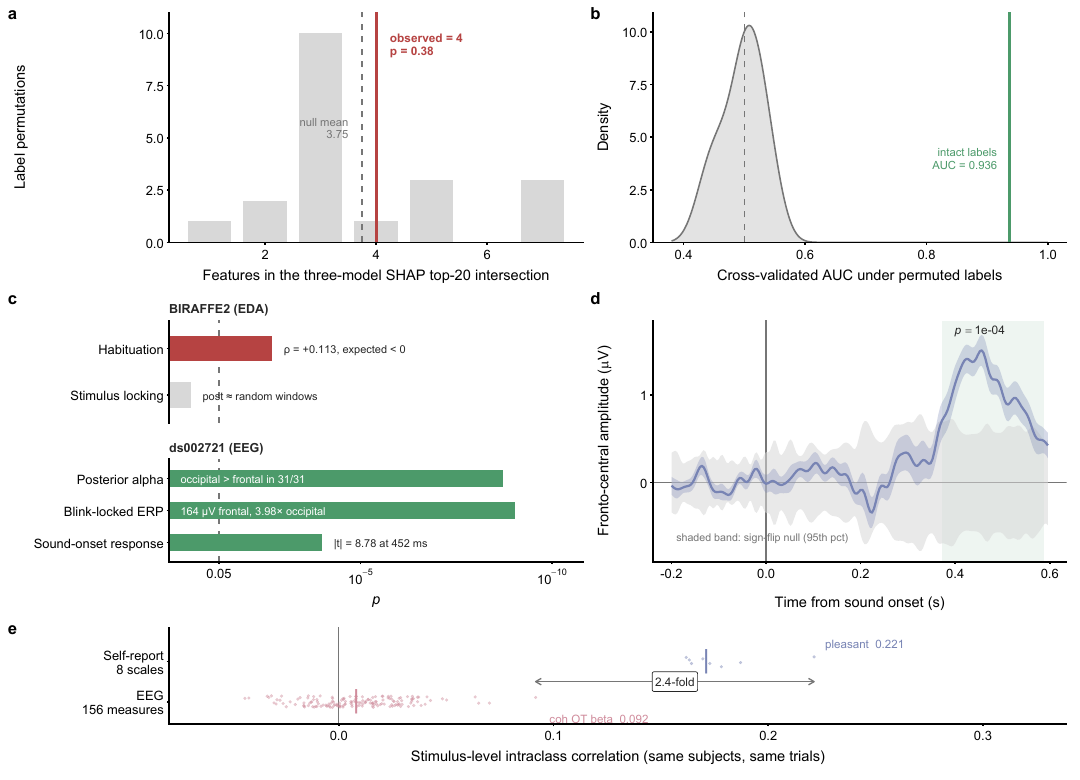}{fig:channel}{\legendfigsix}

\msfigure{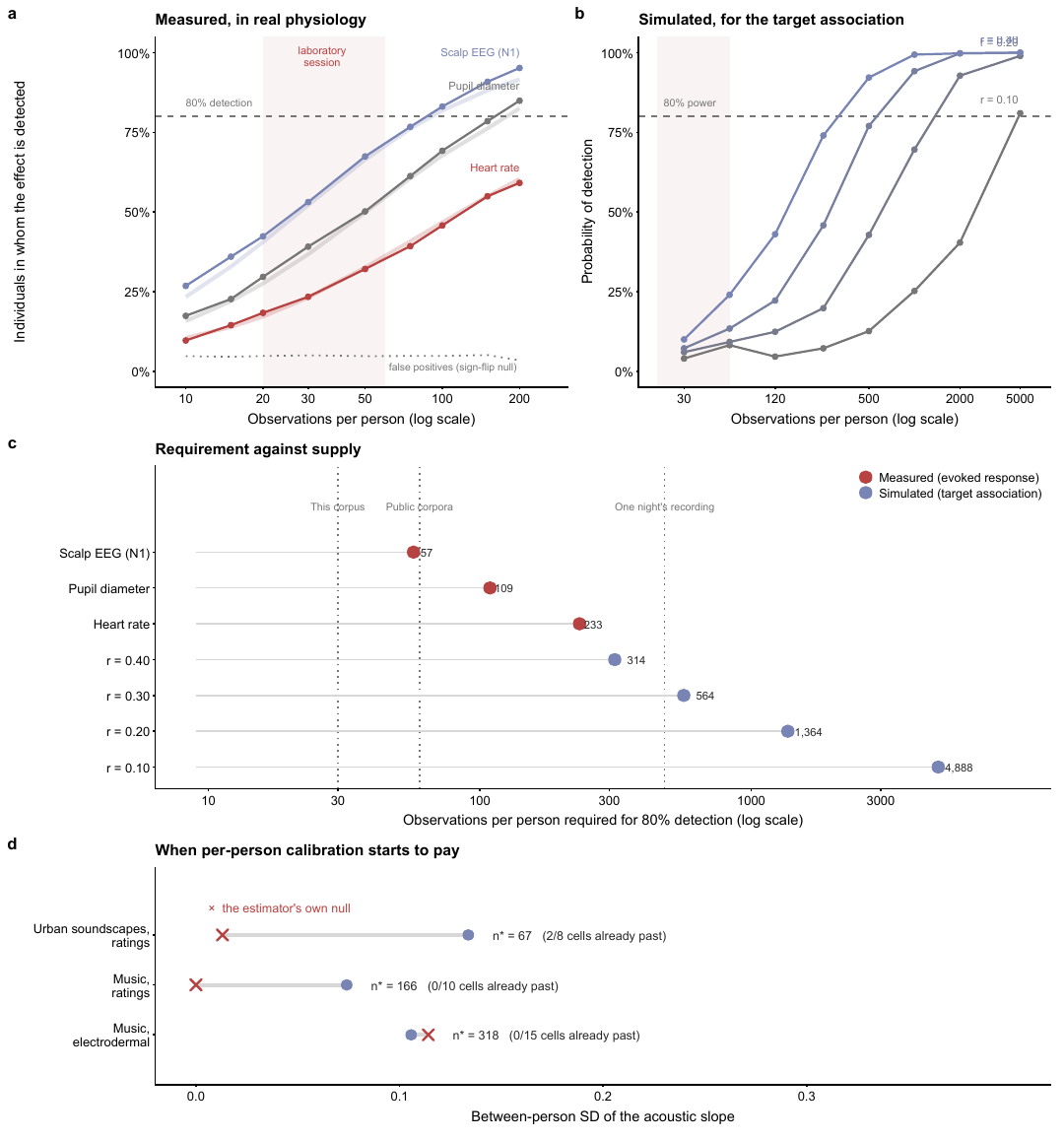}{fig:design}{\legendfigseven}

\msfigure{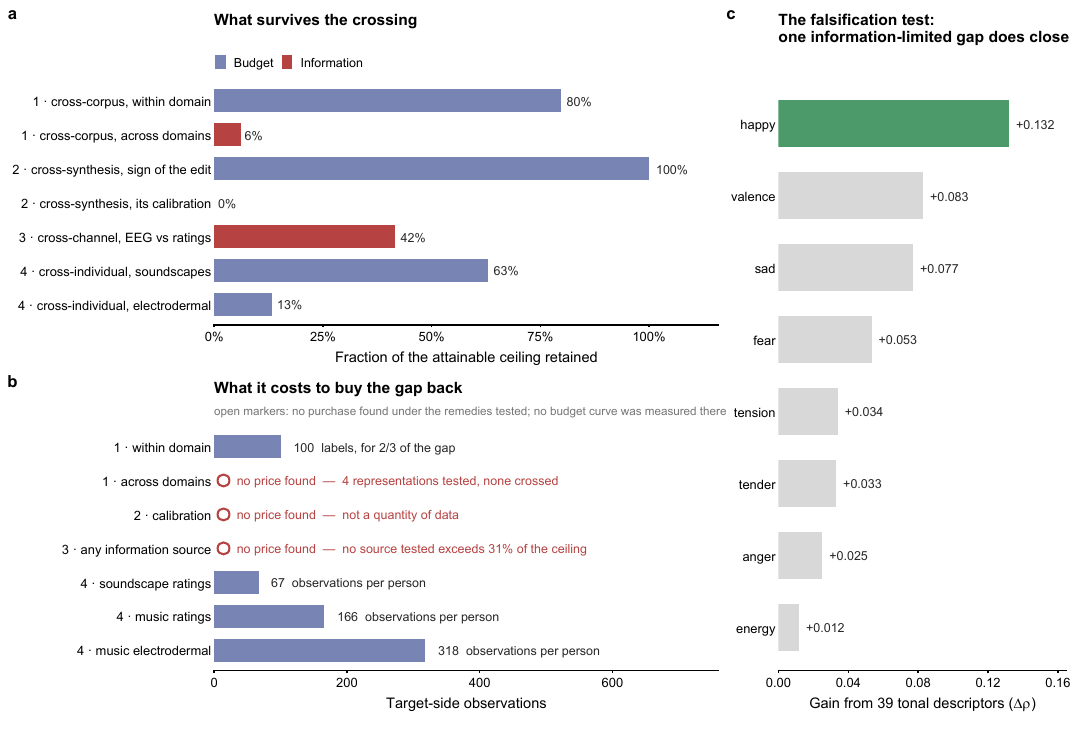}{fig:boundaries}{\legendfigeight}

\end{document}